\documentclass[epsf,epsfig,aps,floatfix,preprint,nofootinbib]{revtex4}
\usepackage{graphics}
\usepackage{graphicx}
\usepackage[english]{babel}
\usepackage{amssymb}
\usepackage[final]{pdfpages}
\usepackage{subfigure}
\usepackage{amsmath}
\usepackage{verbatim}
\usepackage{pstricks}
\usepackage{slashed}
\usepackage[normalem]{ulem}
\usepackage{hyperref}
\usepackage{color}

\newcommand{\nc}{\newcommand}
\nc{\beq}{\begin{equation}}
\nc{\eeq}{\end{equation}}
\nc{\barray}{\begin{eqnarray}}
\nc{\earray}{\end{eqnarray}}
\nc{\barrayn}{\begin{eqnarray*}}
\nc{\earrayn}{\end{eqnarray*}}
\nc{\bcenter}{\begin{center}}
\nc{\ecenter}{\end{center}}
\nc{\ket}[1]{| #1 \rangle}
\nc{\bra}[1]{\langle #1 |}
\nc{\0}{\ket{0}}
\nc{\mc}{\mathcal}
\nc{\er}[1]{(\ref{eq:#1})}
\nc{\onehalf}{\frac{1}{2}}
\nc{\partialbar}{\bar{\partial}}
\nc{\psit}{\widetilde{\psi}}
\nc{\Tr}{\mbox{Tr}}
\nc{\ev}{\;\mathrm{eV}}
\nc{\mev}{\;\mathrm{MeV}}
\nc{\gev}{\;\mathrm{GeV}}

\def\chii0{\chi_i^0}
\def\chij0{\chi_j^0}

\newcommand{\bi}{\begin{itemize}}
\newcommand{\ei}{\end{itemize}}

\newcommand{\gsim}{\lower.7ex\hbox{$\;\stackrel{\textstyle>}{\sim}\;$}}
\newcommand{\lsim}{\lower.7ex\hbox{$\;\stackrel{\textstyle<}{\sim}\;$}}

\def\afb {{A_{FB}}}

\begin{document}

\preprint{MCTP-11-13}
\title{On Models of New Physics for the Tevatron Top $A_{FB}$}
\author{Moira I. Gresham, Ian-Woo Kim and Kathryn M. Zurek}
\affiliation{Michigan Center for Theoretical Physics, Department of Physics, University of Michigan, Ann Arbor, MI 48109}
\date{\today}
\begin{abstract}

CDF has observed a top forward-backward asymmetry discrepant with the Standard Model prediction at $3.4\sigma$.  We analyze models that could generate the asymmetry, including flavor-violating $W'$s, horizontal $Z'_H$s, triplet and sextet diquarks, and axigluons. We consider the detailed predictions of these models for the invariant mass and rapidity distributions of the asymmetry at the parton level, comparing against the unfolded parton-level CDF results.  
While all models can reproduce the asymmetry with the appropriate choice of mass and couplings, it appears at first examination that the extracted parton-level invariant mass distribution for all models are in conflict with Tevatron observations.  We show on closer examination, however, that $t\bar{t}$ events in $Z'_H$ and $W'$ models have considerably lower selection efficiencies in high invariant mass bins as compared to the Standard Model, so that $W'$, $Z'_H$, and axigluon models can generate the observed asymmetry while being consistent with the total cross-section and invariant mass spectrum.  Triplet and sextet models have greater difficulty producing the observed asymmetry while remaining consistent with the total cross-section and invariant mass distribution.  To more directly match the models and the CDF results, we proceed to decay and reconstruct the tops, comparing our results against the ``raw'' CDF asymmetry and invariant mass distributions.   We find that the models that successfully generate the corrected CDF asymmetry at the parton level reproduce very well the more finely binned uncorrected asymmetry.  Finally, we discuss the early LHC reach for discovery of these models, based on our previous analysis \cite{Gresham:2011dg}.

\end{abstract}
\maketitle

\tableofcontents

\section{Introduction}

Recently, there has been an apparent anomaly in the top sector: the observation by the CDF experiment of a top forward-backward asymmetry ($\afb$)~\cite{Aaltonen:2011kc}, where the forward-backward asymmetry in a particular invariant mass bin, $M_{t\bar{t},i}$, is defined by
\begin{equation}
A^{t\bar{t}}(M_{t\bar{t},i}) = \frac{N(\Delta y>0,M_{t\bar{t},i})-N(\Delta y<0,M_{t\bar{t},i})}{N(\Delta y>0,M_{t\bar{t},i})+N(\Delta y<0,M_{t\bar{t},i})},
\label{AFBdef}
\end{equation}    
with $\Delta y$ the rapidity difference between a top and an anti-top. The recent CDF anlaysis shows  $\afb = 0.475 \pm 0.114$ for $M_{t\bar{t}} > 450$ GeV~\cite{Aaltonen:2011kc}, while the Next-to-Leading Order (NLO) Standard Model (SM) predicts much lower values $0.088 \pm 0.013$~\cite{Kuhn:1998jr,Kuhn:1998kw,Bowen:2005ap,Almeida:2008ug}, corresponding to a 3.4$\sigma$ deviation.  The D0 collaboration also observes a larger than predicted asymmetry~\cite{Abazov:2007qb}.  

Since the SM prediction for the top-pair production cross section is in relatively good agreement with observation, a new physics model must generate the large forward-backward asymmetry without disturbing the total cross-section or observed invariant mass spectrum of $t\bar{t}$ production. In order to do this, many models assume an additional tree-level contribution from the exchange of a new particle in a way that maximizes the effect on the forward-backward asymmetry while minimizing the effect on the overall production cross section. Models proposed thus far in the literature that generate the observed Tevatron $\afb$ at tree level while not grossly disrupting the top pair production cross section fall into two categories according to the nature of the new particle exchange: (i) $s$-channel exchange of vector mediators with axial couplings ({\it e.g.} axigluon models)~\cite{Sehgal:1987wi,Bagger:1987fz,Ferrario:2009bz,Frampton:2009rk,Chivukula:2010fk,Djouadi:2009nb,Bauer:2010iq,Alvarez:2010js,Chen:2010hm, Delaunay:2011vv, Bai:2011ed, Zerwekh:2011wf,Barreto:2011au} or (ii) $t$-channel exchange of flavor-violating mediators~\cite{Jung:2009jz,Cheung:2009ch,Shu:2009xf,Dorsner:2009mq,Arhrib:2009hu,Barger:2010mw, Gupta:2010wt, Xiao:2010hm, Gupta:2010wx,Cheung:2011qa,Cao:2011ew,Berger:2011ua,Barger:2011ih,Grinstein:2011yv, Patel:2011eh}.  Comparative studies of these models have also been carried out \cite{Jung:2009pi,Cao:2009uz,Cao:2010zb,Jung:2010yn,Choudhury:2010cd,Jung:2010ri,Blum:2011up,Ahrens:2011mw, Craig:2011an, Delaunay:2011gv,Foot:2011xu}.   The $s$-channel mediators tend to have maximally axial couplings, while the $t$-channel mediators tend to be maximally flavor violating, connecting a light quark to the top quark. Recently, it has been pointed out that such maximal flavor violation can also explain anomalies in the $B_s$ and $B_d$ systems when the $b$-quark is coupled as well~\cite{Shelton:2011hq,note}.

The recent CDF analysis \cite{Aaltonen:2011kc} greatly extends the experimental information about the asymmetry. Due to much improved statistics of top-pair production at the Tevatron,  the analysis shows event distributions in more detail. In particular, CDF has now collected enough data to give the forward-backward asymmetry with respect to the invariant mass, $M_{t\bar{t}}$, and the rapidity, $y$, of reconstructed tops, and to compare these distributions for various data subsets, such as the 4- and 5-jet samples. This alows us to reassess the viability of previously suggested theoretical models that address the $A_{FB}$ anomaly.  For example,  in the high invariant mass bin, $M_{t\bar{t}}>$ 450 GeV, the asymmetry is very large, $\sim 50\%$, and hence some models that marginally explain the previous $A_{FB}$ will be challenged.   Diquark models, for example, have great difficulty in producing a large enough asymmetry without generating an overly large correction to the total production cross-section.  

In addition, although the top forward-backward asymmetry and related variables are straight-forwardly defined observables, the interpretation of experimental observation to confirm or falsify models must be done with care due to the indirect nature of top quark identification. In spite of the effort of the CDF group to show an ``unfolded'' $A_{FB}$ for direct comparison with theory predictions, a broad coverage of models is needed to justify its model independence, especially where selection effects can come into play. Therefore, it is necessary to compare direct experimental signatures to expectations within a wide variety of models.

The purpose of this paper is to give a comprehensive analysis of a variety of models that have been proposed in the literature to explain the $\afb$ anomaly in order to (1) reassess the viability of such models given the new experimental data and (2) investigate subtleties associated with unfolding the ``raw'' $A_{FB}$ to a parton level $A_{FB}$ for comparison to models.\footnote{We use ``raw'' to refer to the measurement uncorrected for detector effects, but with background subtracted, as presented in the CDF paper.}  In order to do the latter, we must decay the tops, sending the events through a detector simulation, and reconstruct the tops utilizing a $\chi^2$-based top reconstruction algorithm.   Because, unlike previous studies, we do fully reconstruct the tops and simulate detector effects, we are able to make comparisons to the raw CDF asymmetries.

Comparing our results against the raw asymmetries and invariant mass spectra also enables us to assess the model dependence of event selection efficiencies.  We find that these efficiencies can dramatically affect the extracted parton level total and differential cross-sections.\footnote{We use ``parton level'' as in the CDF $A_{FB}$ paper, where the term refers to de-convolving event selection efficiencies, detector efficiencies, jet algorithms, background, etc., from the underlying physics \cite{Aaltonen:2011kc}.}  For example, $Z'_H$ and $W'$ models that produce large asymmetries have event selection efficiencies in the high invariant mass bins that are about half or less than what a Standard Model $t\bar{t}$ Monte Carlo would predict.  This must be taken into account when comparing a model against parton level results.   For example, while the parton level invariant mass spectra of $Z'_H$ and $W'$ models appear badly in conflict with the reconstructed invariant mass spectra from CDF at high invariant mass, we find that once the selection efficiencies are taken into account, the agreement between the data and model is good.

This paper is complementary to our earlier paper in which we examined the reach of the LHC at 7 TeV to discover the mediators of $t$-channel physics generating the Tevatron asymmetry.  We found there that a top-flavor violating state light enough and with large enough couplings to generate the asymmetry will be rapidly discoverable at the LHC via a search for $tj$ resonances in $t\bar{t}j$ events \cite{Gresham:2011dg}.  

The outline of this paper is as follows.  In the next section we summarize and  discuss the qualitative features of the classes of models that have been shown to be capable of generating the Tevatron top $A_{FB}$. 
In the following section, we first show our parton-level comparison of various models to the unfolded Tevatron results.  Then we show detector-level comparison utilizing a reconstructed top sample.  Next we present a lepton asymmetry from fully leptonic $t\bar{t}$ events, which has recently been discussed in a CDF note \cite{CDFLeptons}.  Lastly, we discuss the LHC reach for discovering such states, based on the analysis of \cite{Gresham:2011dg}.

\section{Models}

\begin{figure}
\centering
\subfigure[~s-channel $q\bar{q}$]{\includegraphics[width=0.2\textwidth]{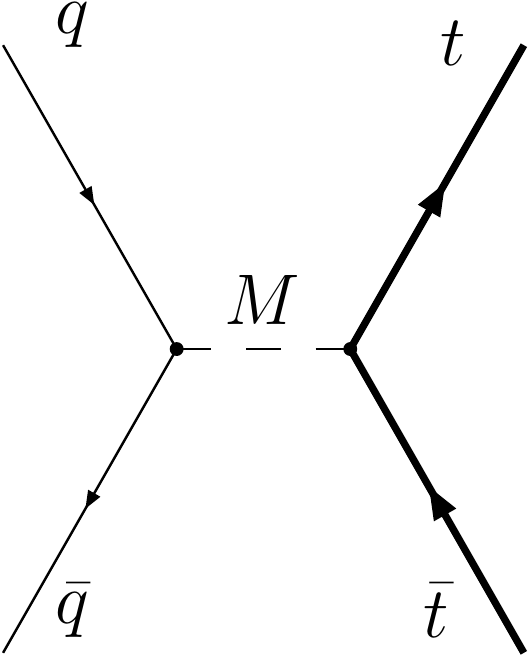}}\qquad
\subfigure[~s-channel $gg$]{\includegraphics[width=0.2\textwidth]{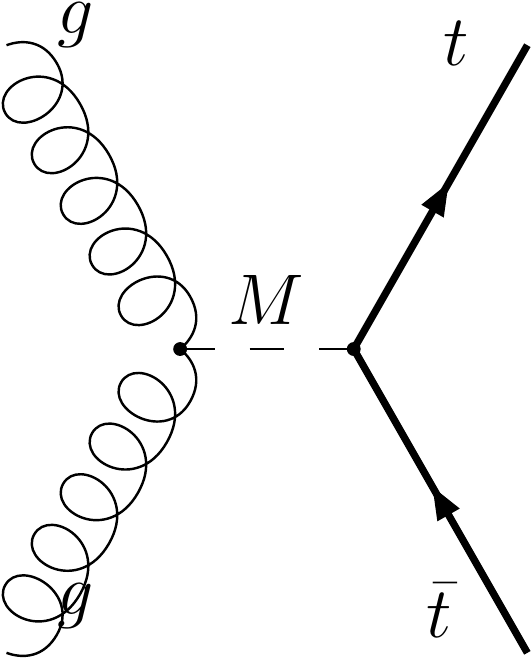}}~~~
\subfigure[~t-channel $q\bar{q}$]{\includegraphics[width=0.2\textwidth]{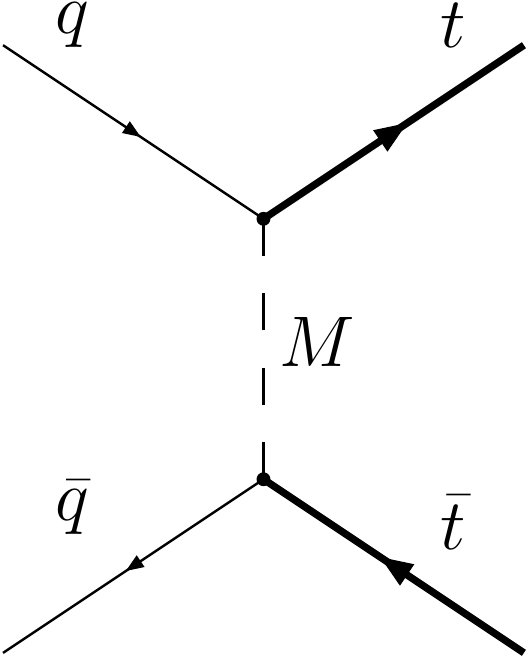}}~~~
\caption[$t \bar{t}$ production diagrams.]{Tree level $t \bar{t}$ production diagram with mediator $M$ exchange.}
\label{feynmandiagram}
\end{figure}

The Leading Order (LO) SM tree-level amplitude for $t\bar{t}$ production does not generate a forward-backward asymmetry. In the SM, a small positive  top forward-backward asymmetry is generated through interference between a one-loop box diagram and a LO tree level diagram, $A_{FB} (M_{t\bar{t}} < 450 {\rm GeV}) = 0.040 \pm 0.006$, $A_{FB} (M_{t\bar{t}} > 450~{\rm GeV}) = 0.088\pm 0.013$.\footnote{ Interference between initial state gluon radiation and final state gluon radiation makes a very small negative contribution to the asymmetry.}  
Since the SM contribution is generated at NLO, if there is an additional LO tree-level contribution from new physics, it can easily dominate.

Such LO diagrams are of the form of those in Fig.~(\ref{feynmandiagram}). They can be either $s$-channel (Fig.~(\ref{feynmandiagram}a) and (\ref{feynmandiagram}b)) or $t$-channel (Fig.~\ref{feynmandiagram}c). $s$-channel mediators couple directly to light flavors and gluons, and therefore the mediator masses must be large enough to evade dijet resonance search constraints \cite{Chivukula:2010fk,Bai:2011ed}. To maximize the contribution to $A_{FB}$, such a model must have a big axial coupling.

On the other hand, $t$-channel models should have large flavor violation between the light and the top generations, as can be seen in Fig.~(\ref{feynmandiagram}c). Large flavor violation 
is experimentally allowed even for low mass mediators, $M$, as long as new couplings between light generations and left-handed quarks is suppressed; then strong limits on flavor violation and from dijet resonance searches are avoided. Additionally, the same-sign top signature search limit prefers $M$ to be a non-self-conjugate state \cite{Jung:2009jz}.  Therefore, ordinary $Z'$ models run into difficulty. Here,  to avoid same-sign top constraints, we consider horizontal $Z_H'$s with flavor charge. Color exotic states and $W'$s can also satisfy the requirement. 

In the following sections, we summarize the defining Lagrangian of $t$-channel $W'$, $Z_H'$, triplet scalar, sextet scalar and $s$-channel axigluon models and present the tree-level differential cross sections, ${d \sigma(q \bar{q} \rightarrow t \bar{t}) \over d {\cos \theta}}$.
 

\subsection{Flavor-Changing $W'$, $Z'$} 

The Lagrangian for a flavor-violating $Z'$ interaction is 
\beq\label{eq:FVZprimeLagrangian}
{\cal L} = \frac{1}{\sqrt{2}}\bar{t} \gamma^\mu (g_L P_L +g_R P_R) u Z'_\mu + \mbox{h.c.},
\eeq
giving rise to a scattering cross-section  
 \beq
 \frac{d \sigma}{d\cos\theta} =\frac{\beta}{32 \pi \hat{s}} \left({\cal A}_{SM} + {\cal A}_{int} + {\cal A}_{sq}\right),
 \eeq
 where
 \beq
 {\cal A}_{SM} = \frac{2 g_s^4}{9}(1+c_\theta^2 + \frac{4 m_t^2}{\hat{s}}),
 \eeq
 with $c_\theta = \beta \cos \theta$ and $\beta = \sqrt{1-4 m_t^2/\hat{s}}$.  The new physics contributions are \cite{Cao:2010zb}
\begin{eqnarray}
{\cal A}_{int}  =  \frac{2 g_s^2 }{9} \frac{(g_L^2+g_R^2)}{\hat{s} \hat{t}_{Z'}}\left[2 \hat{u}_t^2+2 \hat{s} m_t^2+\frac{m_t^2}{m_{Z'}^2}(\hat{t}_t^2+\hat{s} m_t^2)\right],
\label{Z'int}
 \end{eqnarray}
 \begin{eqnarray}
{\cal A}_{sq}  =   \frac{1}{2 \hat{t}_{Z'}^2}&&\left[(g_L^4+g_R^4)\hat{u}_t^2+2g_L^2g_R^2 \hat{s} (\hat{s}-2 m_t^2) +\frac{m_t^4}{4 m_{Z'}^4}(g_L^2+g_R^2)^2(\hat{t}_{Z'}^2+4 \hat{s} m_{Z'}^2)\right],
\label{Z'sq}
 \end{eqnarray}
 with $\hat{t}_i \equiv \hat{t} - m_i^2$ and $\hat{u}_i \equiv \hat{u} - m_i^2$. The Mandelstam variables are related to the scattering angles via $\hat{t} = -\hat{s}(1-c_\theta)/2 + m_t^2$ and $\hat{u} = -\hat{s}(1+c_\theta)/2 + m_t^2$. Note that the Lagrangian has been defined with a $\sqrt{2}$ with respect to some other conventions in the literature. Similar expressions hold for the flavor-violating $W'$ via the interaction Lagrangian
\beq\label{eq:FVWprimeLagrangian}
{\cal L} = \frac{1}{\sqrt{2}}\bar{d} \gamma^\mu (g_L P_L +g _R P_R) t {W'}_\mu + \mbox{h.c.}
\eeq


\subsection{Color Triplets and Sextets} 

The quantum numbers of the color triplet and sextet are
\beq
(\bar{3},1)_{4/3}~~~~(6,1)_{4/3},
\eeq
and their interactions with up and top quarks is given by
\beq\label{eq:TripSextetLagrangian}
{\cal L}_\phi = \phi^a \bar{t^c}T_r^a(g_L P_L + g_R P_R) u.
\eeq
This gives rise to a scattering cross-section \cite{Shu:2009xf}
\beq
{\cal A}_{int} + {\cal A}_{sq} = \frac{2 g^2 g_S^2 C_{(0)}}{9}\frac{\hat{u}_t^2 + \hat{s} m_t^2}{\hat{s} \hat{u}_\phi}+\frac{g^4 C_{(2)}}{9}\frac{\hat{u}_t^2}{\hat{u}_\phi^2}
\label{tripintsq}
\eeq
where $C_{(0)} = 1 (-1)$ for triplets (sextets) \cite{Shu:2009xf,Arhrib:2009hu} is a color factor that comes from the interference of new $t$-channel physics with the $s$-channel gluon. The color factor $C_{(2)}$ comes from the squared new $t$-channel physics term and is equal to $C_{(2)} = 3/2$ for sextets  and $C_{(2)} = 3/4$ for triplets.  We have also defined
$
g \equiv \sqrt{(g_L^2 + g_R^2)/2}.
$


\subsection{Color Octet} 

The exotic gluon couples to light quarks through
\begin{eqnarray}\label{eq:ExoticGluonLagrangian}
{\cal L}_{axi} = 
				g_s \left(\bar{q} \, T^A \gamma^\mu(g_L^q P_L+g_R^q P_R) q + \bar{t} \, T^A \gamma^\mu(g_L^t P_L+g_R^t P_R) t \right) {G'}^A_\mu. 
\end{eqnarray}
Note the inclusion of the QCD coupling constant, $g_s$, in the interaction. The scattering cross-sections calculated through these interactions are \cite{Cao:2010zb} 
 \begin{eqnarray}
{\cal A}_{int} & = & \frac{g_s^4}{9}\frac{\hat{s}(\hat{s}-m_{G'}^2)}{(\hat{s}-m_{G'}^2)^2+m_{G'}^2\Gamma_{G'}^2}(g^q_L+g^q_R)(g^t_L+g^t_R) \nonumber \\
&&\left[(2 - \beta^2)+2\frac{(g_L^q-g_R^q)(g_L^t-g_R^t)}{(g_L^q+g_R^q)(g_L^t+g_R^t)} c_\theta +c_\theta^2\right],
 \end{eqnarray}
 \begin{eqnarray}
{\cal A}_{sq} & = & \frac{g_s^4}{18}\frac{\hat{s}^2}{(\hat{s}-m_{G'}^2)^2+m_{G'}^2\Gamma_{G'}^2}({g_L^q}^2+{g_R^q}^2)({g_L^t}^2+{g_R^t}^2) \nonumber \\
&&\left[1+(1 - \beta^2)\frac{2 g_L^t g_R^t}{{g_L^t}^2+{g_R^t}^2}+2\frac{({g_L^q}^2-{g_R^q}^2)({g_L^t}^2-{g_R^t}^2)}{({g_L^q}^2+{g_R^q}^2)({g_L^t}^2+{g_R^t}^2)}c_\theta+c_\theta^2\right].
 \end{eqnarray}

As does CDF, we consider the case where the couplings of the vector color octets are purely axial, so $g_V^q = (g_R^q + g_L^q)/2 = 0$ and $g_V^t = (g_L^t + g_R^t)/2 = 0$, and where the axial coupling of the boson to light quarks is positive and opposite the coupling of the boson to tops, so $g_A^t = (g_R^t - g_L^t)/2 = - g_A^q = (-g_R^q + g_L^q)/2$. This axigluon case leads to the largest positive contribution to the asymmetry per contribution to the total cross-section.

\section{Parton Level Tevatron Top Forward-Backward Asymmetry}

We now simulate the models described in the previous section, and we will use the formulae presented there to discuss the qualitative features.

\begin{figure}
\includegraphics[width = 0.95\textwidth]{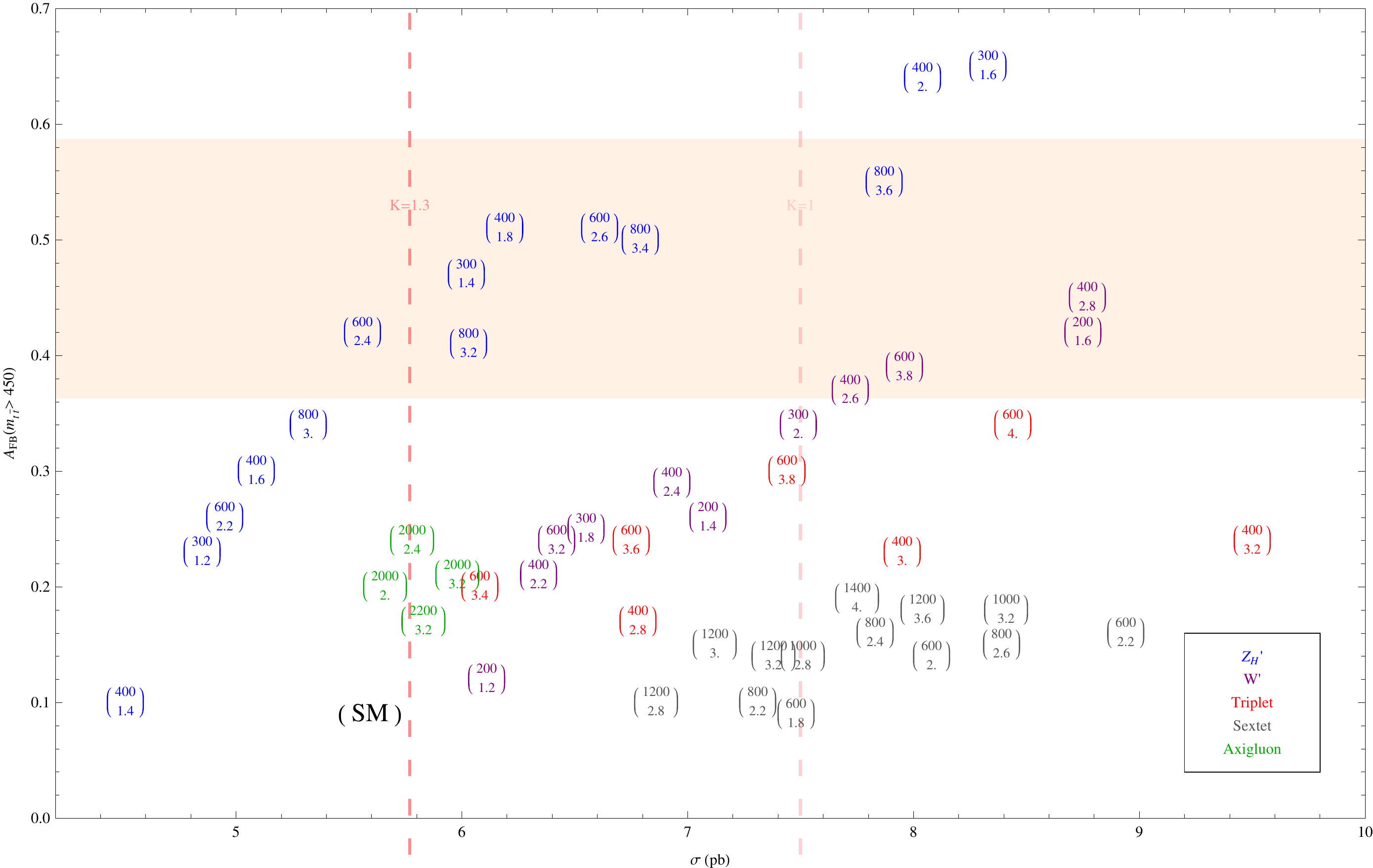}
\caption[Scatter plot of $A_{FB}(m_{t \bar{t}})$ versus LO cross-section.]{Parton level forward-backward asymmetry for events with $t \bar{t}$ invariant mass greater than $450$ GeV versus {\em leading-order} cross-section in picobarns for various models.  The mass (in GeV) and coupling of the mediator are indicated by ${ \text{mass} \choose \text{coupling} } $.   A comprehensive scan of models was carried out, and only representative points are shown.  The coupling shown for axigluons is $g_A^q = -g_A^t$ (supposing $g_V^q = g_V^t = 0$) while the coupling shown for $Z_H'$ and $W'$ models is $g_R$ (assuming $g_L = 0$) and the quoted couplings for triplets and sextets is $g = \sqrt{(g_L^2 + g_R^2)/2}$.  For comparison against observations, the horizontal shaded area lies in the $\pm 1\sigma$ region of the measured (parton-level) value of $A_{FB}(M_{t \bar{t}} > 450~\text{GeV})$. The vertical lines lie at the central value of the CDF $t\bar{t}$ production cross-section ($7.5 \pm 0.48$ pb), divided by a K-factor of 1.3 or 1. The SM marker lies at the value of the LO Standard Model cross-section and the NLO value for the SM forward-backward asymmetry.  Note that care must be taken when comparing the new physics cross-sections against the Standard Model cross-section, as the selection efficiencies for the new physics models can be lower. This is discussed in more detail in the main text.}\label{modelScatterPlot}
\end{figure}

We begin by analyzing models for top asymmetry generation at the parton level.   ``Parton level'' as used in the CDF $A_{FB}$ paper refers to de-convolving event selection efficiencies, detector efficiencies, jet algorithms, background, etc., from the underlying physics \cite{Aaltonen:2011kc}. Thus for the parton level analysis, we simulate and compare our results after showering in {\tt PYTHIA} but before folding in detector effects.  To do this, we use {\tt MadGraph/MadEvent} with matrix element / parton shower (ME/PS) matching in the MLM scheme, which was implemented in the {\tt MadGraph/MadEvent} package using {\tt PYTHIA}.  The events were generated using a fixed renormalization scale and factorization scale of 200 GeV.\footnote{The current version of {\tt MadGraph} uses $\alpha_s(m_Z) = 0.13$, which is substantially larger than the current measured value of $\alpha_s = 0.118$. Therefore our choice of the renormalization scale is effectively lower than the nominal scale, assuming $\alpha_s(m_Z) = 0.118$. Our choice reproduces well the known theoretical LO Standard Model value for the $t \bar{t}$ cross-section at Tevatron. }   {\tt MadGraph5 v0.6.1 / MadEvent 4.4.44} with {\tt QCUT} = 30 and {\tt xqcut} = 20 was used to generate all signal and standard model events. 
Showering and matching are done in order to improve accuracy, including the effects of single mediator production in inclusive $t \bar{t}$ events.

\begin{figure}
\includegraphics[height = 0.3\textwidth]{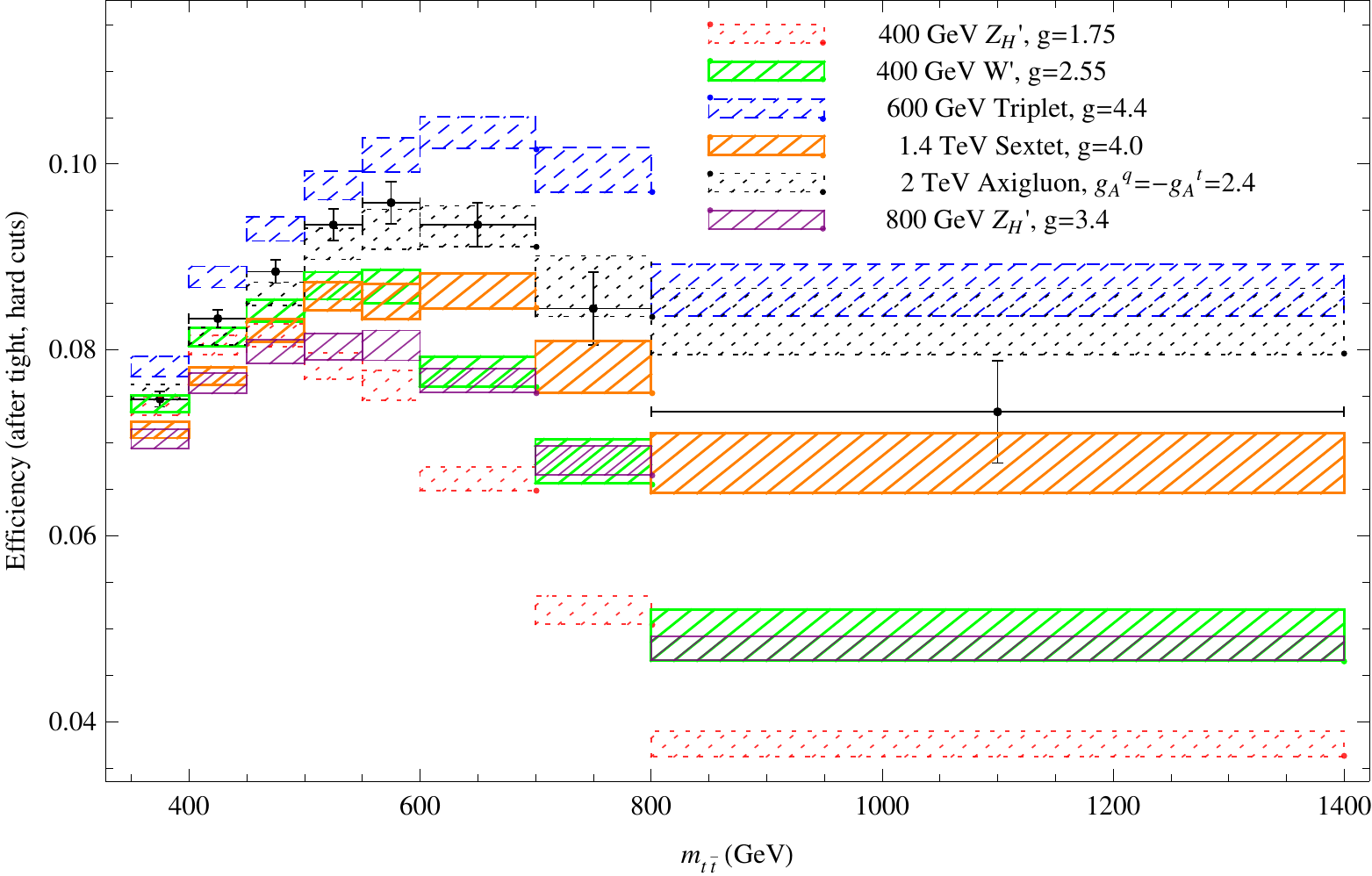}
\includegraphics[height = 0.3\textwidth]{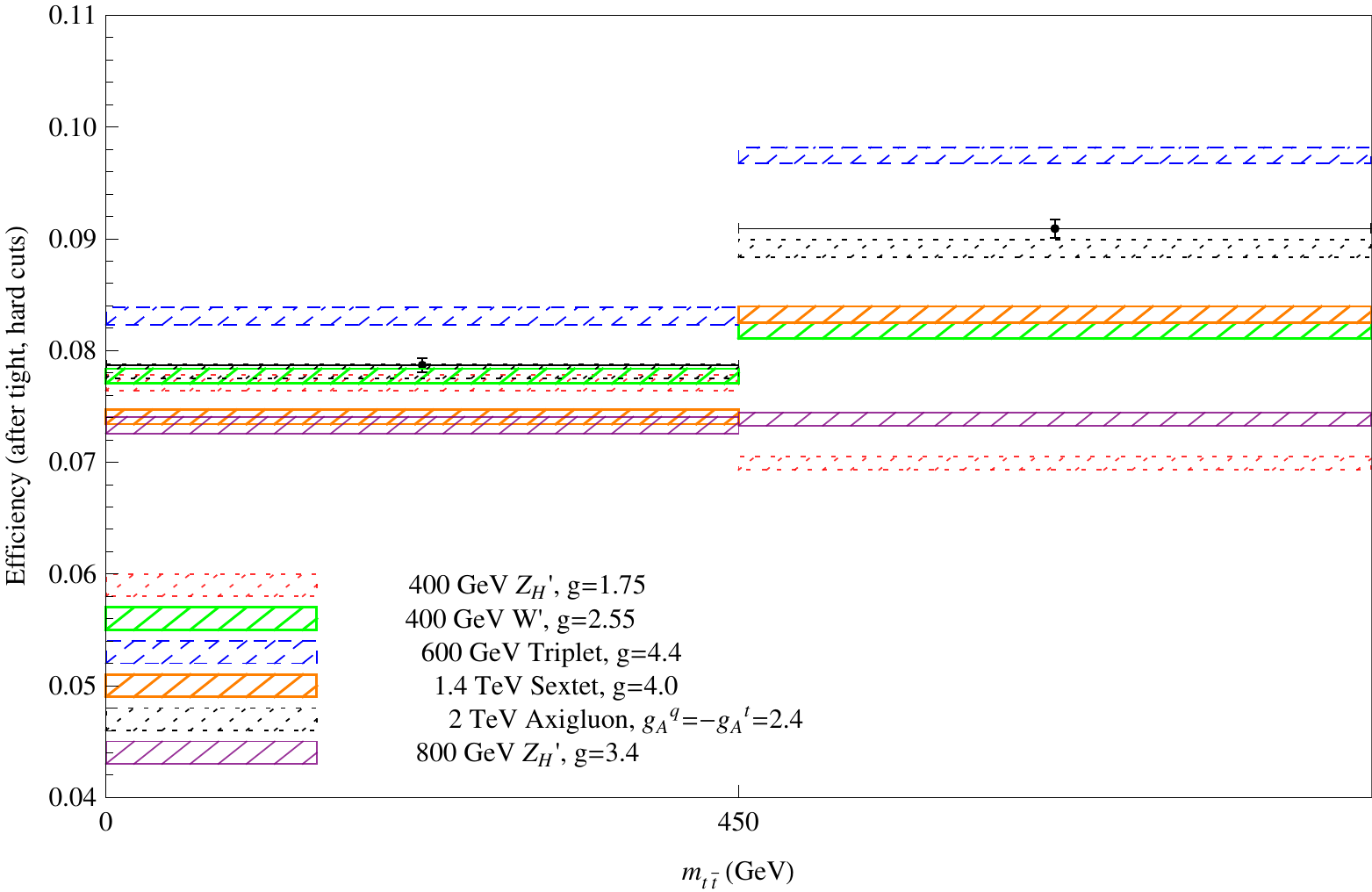}
\caption[Efficiencies by mass bins.]{ Efficiencies after showering events with {\tt PYTHIA} and making the cuts detailed in section~\ref{Sec:reconstructed} on final state leptons and jets, with the hatched regions corresponding to $1\sigma$ errors based on the limited statistics of the sample.  No detector effects have been taken into account here.  Black bars are for LO Standard Model. All samples are LO matched.}\label{efficiencies}
\end{figure}

\subsection{Cross-sections and Efficiencies}

We make a few observations about total cross-sections and invariant mass distributions before delving into a detailed analysis of the asymmetry.  Fig.~(\ref{modelScatterPlot}) shows  the parton level $A^{t \bar{t}}_{FB}$ for events with $m_{t \bar{t}} > 450$ GeV versus total leading order $p \bar{p} \rightarrow t \bar{t}$ + 0 or 1 jet cross-section for a swath of flavor-changing $W'$, $Z'_H$, triplet, sextet models, and axigluon models.  A comprehensive scan of models was carried out, and only representative points are shown.  The horizontal shaded band lies at $\pm 1\sigma$ values of the observed asymmetry in the high invariant mass bin, $M_{t\bar{t}} > 450 \mbox{ GeV}$. The vertical dashed lines correspond to the combined $t \bar{t}$ cross-section from CDF with 4.6 fb$^{-1}$; CDF measures a cross-section of $7.5 \pm 0.31 (\mbox{stat}) \pm 0.34 (\mbox{syst}) \pm 0.15(\mbox{lumi}) $, assuming a top mass of $m_t = 172.5$ GeV \cite{cdftotalxsection}. The predicted next-to-leading-order (NLO) SM cross-section at the value of the top mass we assumed in simulations, $m_t = 174.3$, is about $7.2$ pb \cite{Moch:2008ai}, whereas we find the LO SM cross-section is $5.6$ pb, implying a SM K-factor of about $7.2 / 5.6 \approx 1.3$.  Of course, the NLO corrections to the new physics have not been calculated, so any comparison between the observed cross-section and the $t\bar{t}$ production cross-section is subject to some uncertainty.  We do choose to show in Fig.~(\ref{modelScatterPlot}), however, the central value of the combined CDF $t\bar{t}$ production cross-section (7.5 pb) divided by the SM K-factor when comparing to the leading-order (LO) $t\bar{t}$ production cross-section of SM plus new physics against the observed production cross-section.  From the figure it is clear that, in general, excepting the $Z'_H$ and axigluon models, models with couplings that are small enough to be in accord with the observed cross-section do not produce a large enough asymmetry in the high $t\bar{t}$ invariant mass bin.

\begin{figure}
\begin{center}
\includegraphics[width=0.32\textwidth]{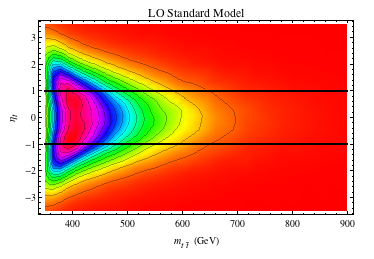}\\
\includegraphics[width=0.95\textwidth]{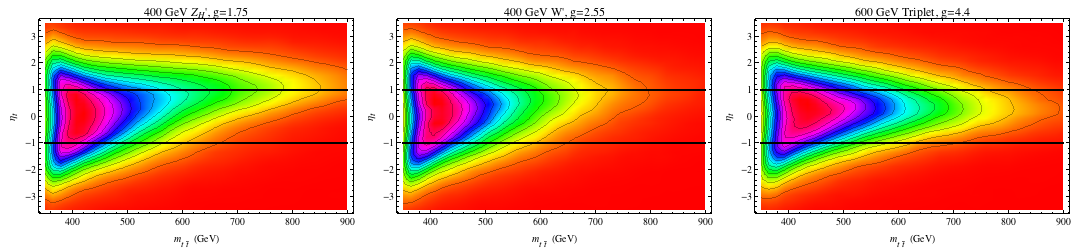}\\
\includegraphics[width=0.95\textwidth]{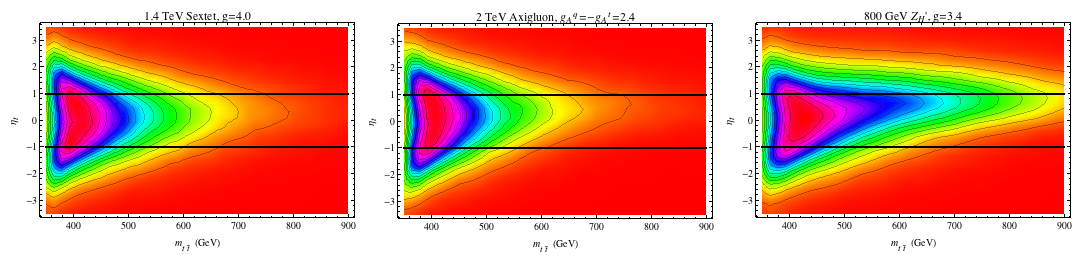}
\caption[Event distributions densities for benchmark models.]{Event distribution densities for benchmark models at the parton level. The vertical axis is top quark pseudo-rapidity, and the horizonal axis it $t \bar{t}$ invariant mass.}\label{contours}
\end{center}
\end{figure}

This statement requires a strong qualification, however, which we investigate in detail below after carrying out a reconstruction of the top samples in these models.  The qualification is that the efficiency for a $t\bar{t}$ event to pass cuts (the same as those used in the CDF analysis and our detector level analysis below, see Sec.~(\ref{Sec:reconstructed})) is strongly model dependent for cases where there is a large asymmetry.  This is shown in Fig.~(\ref{efficiencies}), where we see that the efficiencies to pass cuts (after showering and jet clustering but no detector simulation) is suppressed by more than a factor of two relative to the Standard Model for the 400 GeV $Z'_H$ model shown and by about a factor of $1.5$ for the 800 GeV $Z'_H$ and the $W'$ model. The reason for this becomes clear after examining the distribution of events in top pseudorapidity, $\eta_t$, and and $t \bar{t}$ invariant mass, $m_{t \bar{t}}$, as shown in Fig.~(\ref{contours}). In order to generate a large asymmetry, the $\eta_t$ distribution must be skewed significantly with respect to the SM distribution. To generate a very large asymmetry in the high invariant mass bin, the distribution must be more skewed at high invariant mass. The distributions for the $Z'_H$ and $W'$ models are so skewed at high invariant mass that the peak of the distribution lies close to the $\eta_t = 1$ line. Thus a cut on lepton pseudorapidity of $|\eta| \leq 1$ and jet rapidity of $|\eta| \leq 2$ ends up cutting out a significantly greater fraction of events at high invariant mass than in the SM case.
Importantly, in unfolding the differential $t \bar{t}$ cross-section, assumptions about event selection efficiencies must be made; the assumption is that actual event selection efficiencies do not differ substantially from the Standard Model efficiencies \cite{Aaltonen:2009iz,CDF:thesis}.  This assumption clearly breaks down in the case of the $Z'_H$ and $W'$. We investigate this effect in the reconstructed sample in the next section.  For now, we compare the parton level asymmetries and cross-sections, and note caveats where this effect is important.

\begin{table}
\begin{tabular}{|c || c | c || c | c ||| c | c || c | c |}
\hline
 &  \multicolumn{4}{| c |||}{$m_{t \bar{t}} < 450 $ GeV } 										&  \multicolumn{4}{| c |}{$m_{t \bar{t}} > 450 $ GeV } \\
 \hline
 &  \multicolumn{2}{| c ||}{$ \Delta y < 0 $} &   \multicolumn{2}{| c |||}{$\Delta y > 0$}  	& \multicolumn{2}{| c ||}{$\Delta y < 0 $} &   \multicolumn{2}{| c |}{$\Delta y > 0 $}  \\
\hline
 Model &  \emph{eff.} & $r$ &  \emph{eff.} & $r$  			& \emph{eff.} & $r$ &  \emph{eff.} & $r$ \\
\hline
 SM &0.079& 0.31&0.078& 0.3  		&0.092& 0.2&0.089& 0.19\\
${Z_H'}^1$&0.078& 0.21&0.076& 0.22 	&0.088& 0.15&0.063& 0.42\\
 $W'$&0.079& 0.23&0.077& 0.27		&0.095& 0.16&0.075& 0.34\\
 Triplet&0.084& 0.18&0.083& 0.23		&0.103& 0.2&0.095& 0.39\\
 Sextet&0.075& 0.26&0.073& 0.28		&0.087& 0.19&0.08& 0.27\\
 Axigluon&0.079& 0.26&0.077& 0.31	&0.096& 0.14&0.086& 0.28\\
 ${Z_H'}^2$&0.074& 0.18&0.072& 0.19	&0.089& 0.16&0.069& 0.47\\
\hline
\end{tabular}
\caption{Parton level efficiencies, \emph{eff.}, and bin fractions, $r$, for the Standard Model (SM) and for benchmark models ${Z_H'}^1$  ( $400$ GeV, $g_R = 1.75$), ${Z_H'}^2$  ( $800$ GeV, $g_R = 3.4$),  $W'$ (400 GeV, $g_R = 2.55$), Triplet ($600$ GeV, $g = 4.4$), Sextet ($1.4$ TeV, $g = 4.0$), and Axigluon ($2$ TeV, $g_A^q = - g_A^t = 2.4$). Here $\text{\emph{eff.}} \equiv { \text{\# events in bin after cuts}  \over \text{\# events in bin before cuts}}$ and $r \equiv { \text{\# events in bin after cuts}  \over \text{total \# events after cuts}}$. }\label{coarse bin efficiencies}
\end{table}

Efficiencies can also affect the forward-backward asymmetry. If, for example, the efficiency for an event to pass cuts is lower for events with $\Delta y > 0$ than for events with $\Delta y < 0$, then cuts will wash out the asymmetry. We show the efficiency to pass cuts (after showering and jet clustering but no detector simulation) in four $m_{t \bar{t}}$, $\Delta y$  bins for the SM and for several benchmark models in Table~\ref {coarse bin efficiencies}. The difference between SM efficiencies and new physics model efficiencies is not as great for the coarse $m_{t \bar{t}}$ binning, implying that the effect on the coarse binned asymmetries reported by CDF will not be as great as for the more finely binned invariant mass distribution.  However, we also see from the table that the efficiencies for $\Delta y > 0$ are smaller for some new models, such as the $Z'_H$ and $W'$ than for the SM and the other new models, implying more washout of the asymmetry for the $Z'_H$ and $W'$, which should be compensated.  These effects will be seen when we compare our parton level asymmetries to the reconstructed asymmetries.  The overall point here is that the efficiencies have some effect on unfolding the parton level asymmetries, but they are not as large an effect as on the invariant mass distribution.

\begin{table}
\begin{tabular}{| l |  c  |  c |  c| c| c | c| }
\hline
Model 		&   		\multicolumn{6}{|c|}{ Mass(GeV), coupling, cross-section(pb) }		 								\\
\hline
FV $W'$		&{200, 1.4, 7.1}&{300, 1.8, 6.5}&{400, 2.4, 6.9}&{400, 2.6, 7.7}&{600, 3.4, 6.9}&{600, 3.6, 7.4}\\ 	
\hline
FV $Z_H'$	&{300, 1.4, 6.}&{400, 1.6, 5.1}&{400, 1.8, 6.2}&{600, 2.4, 5.6}&{800, 3.2, 6.}&{800, 3.4, 6.8} \\		
\hline
Triplet		&{400, 3., 7.9}&{400, 3.2, 9.5}&{600, 3.6, 6.7}&{600, 3.8, 7.4}&{600, 4., 8.4} &	\\					
\hline
Sextet		&{600, 2., 8.1}&{800, 2.4, 7.8}&{1000, 3., 8.}&{1200, 3., 7.1}&{1200, 3.4, 7.7}&{1400, 4., 7.7}  \\
\hline
Axigluon		&{2000, 2., 5.7}&{2000, 2.4, 5.8}&{2000, 3.2, 6.}&{2200, 3.2, 5.8}& & \\
\hline
\end{tabular}
\caption{Summary of benchmark models. 
The coupling shown for axigluons is $g_A^q = -g_A^t$ (supposing $g_V^q = g_V^t = 0$) while the coupling shown for $Z_H'$ and $W'$ models is $g_R$ (assuming $g_L = 0$) and the quoted couplings for triplets and sextets is $g = \sqrt{(g_L^2 + g_R^2)/2}$. 
For each model we consider, we include mass, coupling and total leading order matched $t\bar{t} + 0~\text{or}~1~\text{jet}$ production cross-section as calculated with {\tt  MadGraph/MadEvent/Pythia}. The cross-sections above can be compared to the cross-section we obtained for the Standard Model using the same cuts and SM input parameters: 5.6 pb.  No K-factors have been included in these quoted cross-sections.  Note that care must be taken when comparing the new physics cross-sections against the Standard Model cross-section, as the selection efficiencies for the new physics models can be lower. This is discussed in more detail in the main text.}
\label{BenchmarkModels}
\end{table}

\subsection{Parton Level Asymmetries}

We now show the parton level asymmetries for each of the benchmark models. We choose several benchmark masses/couplings for $Z_H'$, $W'$, triplet, sextet, and axigluon models that give rise to large forward-backward asymmetries without  generating too large a total cross-section. These benchmark models are listed in Table~\ref{BenchmarkModels}.

\begin{figure}
\begin{center}
\includegraphics[width=0.45\textwidth]{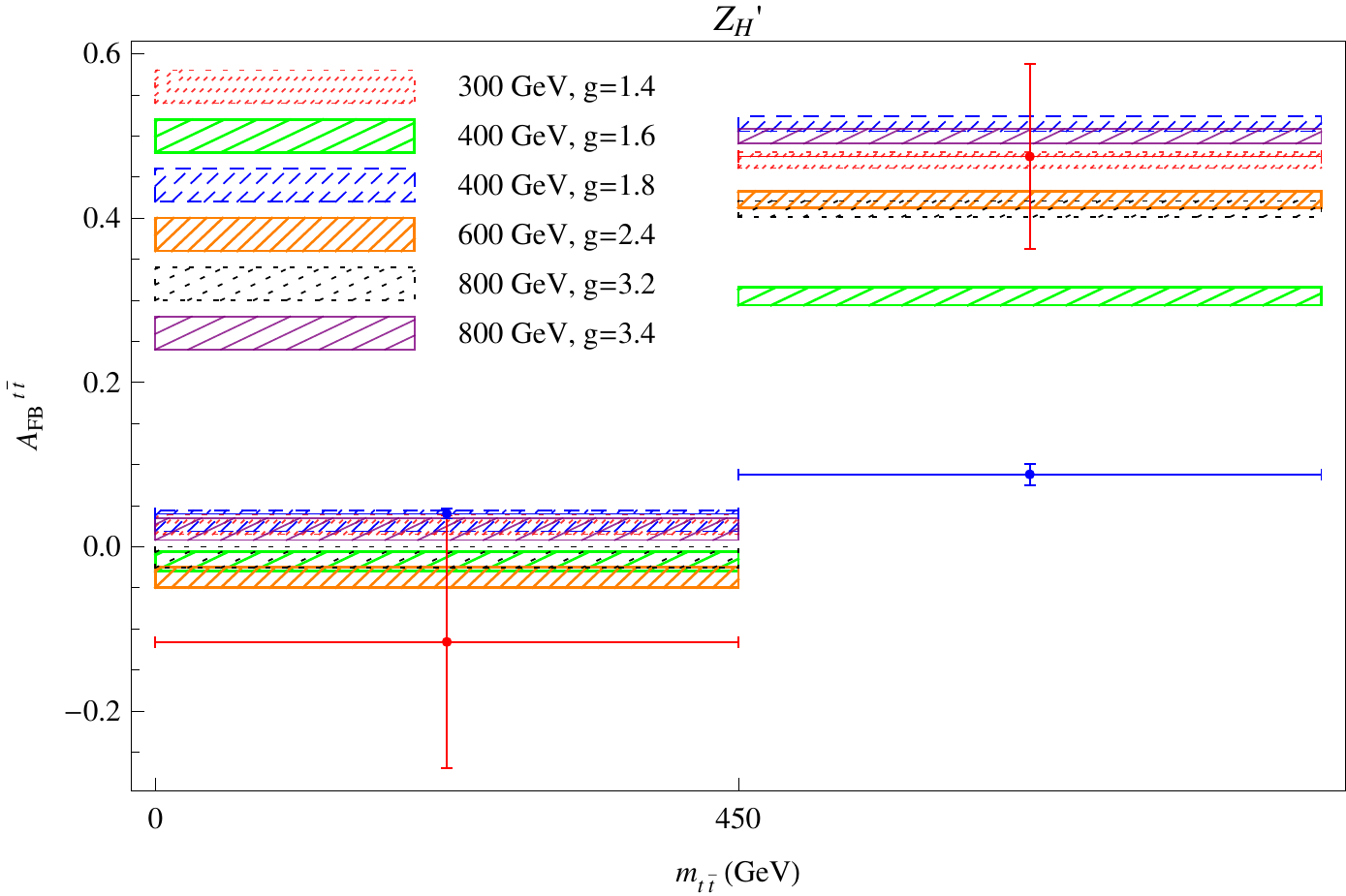}
\includegraphics[width=0.45\textwidth]{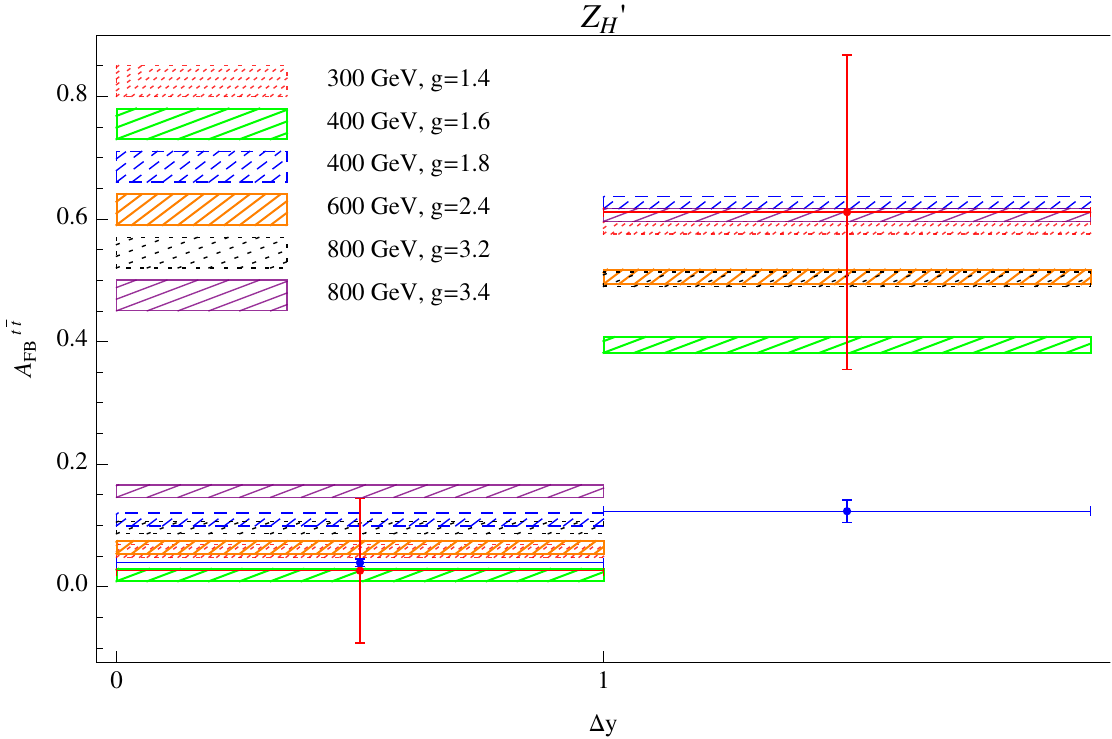}\\
\includegraphics[width=0.45\textwidth]{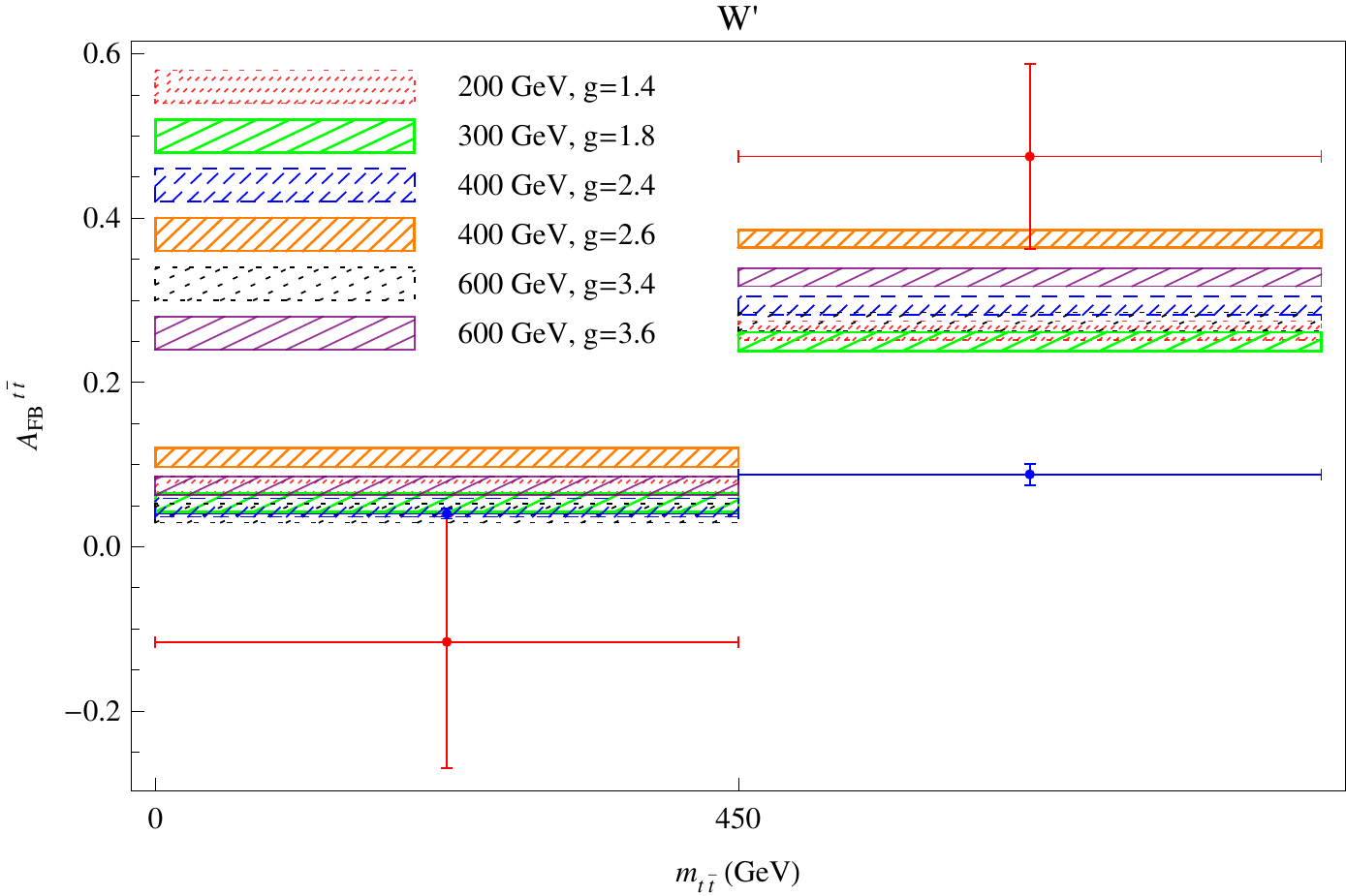}
\includegraphics[width=0.45\textwidth]{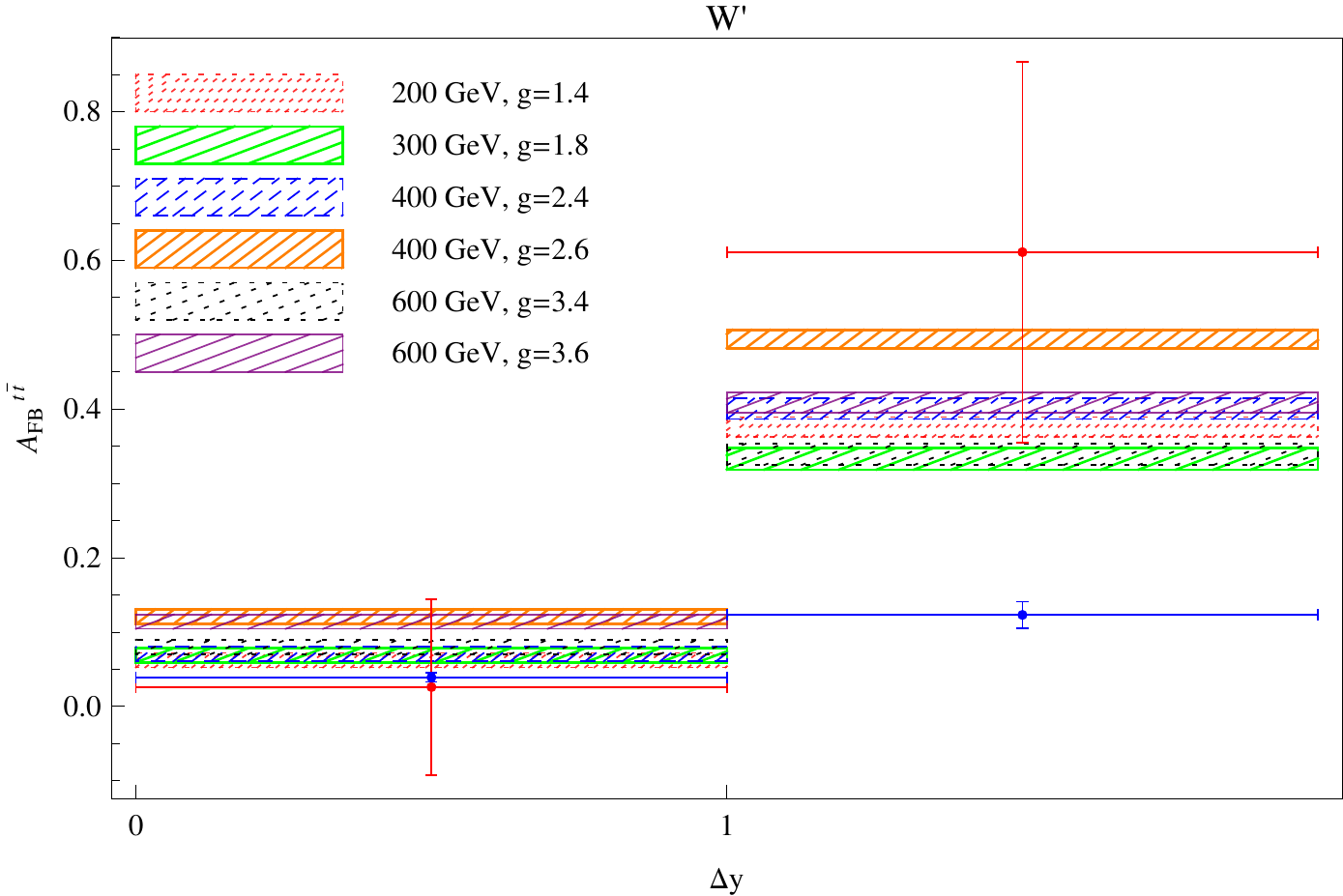}
\end{center}
\caption[Parton level $A_{FB}$ for $Z'_H,~W'$ models.]{$A_{FB}$ for $Z'_H,~W'$ models with couplings as indicated in the legend (with $g=g_R, g_L=0$), with the hatched regions corresponding to $1\sigma$ errors based on the limited statistics of the sample. 
The contribution to $A_{FB}$ includes both $t$-channel $Z'_H,~W'$ exchange and single $Z'_H,~W'$ production.  
The red bars are the CDF observation with $1\sigma$ errors, while the blue bars indicate the NLO SM contribution from \cite{Aaltonen:2011kc}, which has not been included in the LO contribution calculated via {\tt MadGraph} and {\tt PYTHIA}. The last bin includes all events with $m_{t \bar{t}} > 450$ GeV and $|\Delta y| > 1$, respectively.}
\label{ZpWpMttDely}
\end{figure}

We show in Fig.~(\ref{ZpWpMttDely}) the forward-backward asymmetry in $M_{t\bar{t}}$ and $\Delta y$, comparing the reconstructed parton level asymmetry of the Tevatron to our simulated matched sample.  The hatched regions correspond to the $1\sigma$ errors based on the limited statistics of the sample. 
We can see that the horizontal $Z'$ and $W'$ can give good fits in both the high and low invariant mass bins, and in low and high $\Delta y$.   
At least for the two bin case, the models also appear to give the correct shape as a function of $M_{t\bar{t}}$ and $\Delta y$.

\begin{figure}
\begin{center}
\includegraphics[width=0.45\textwidth]{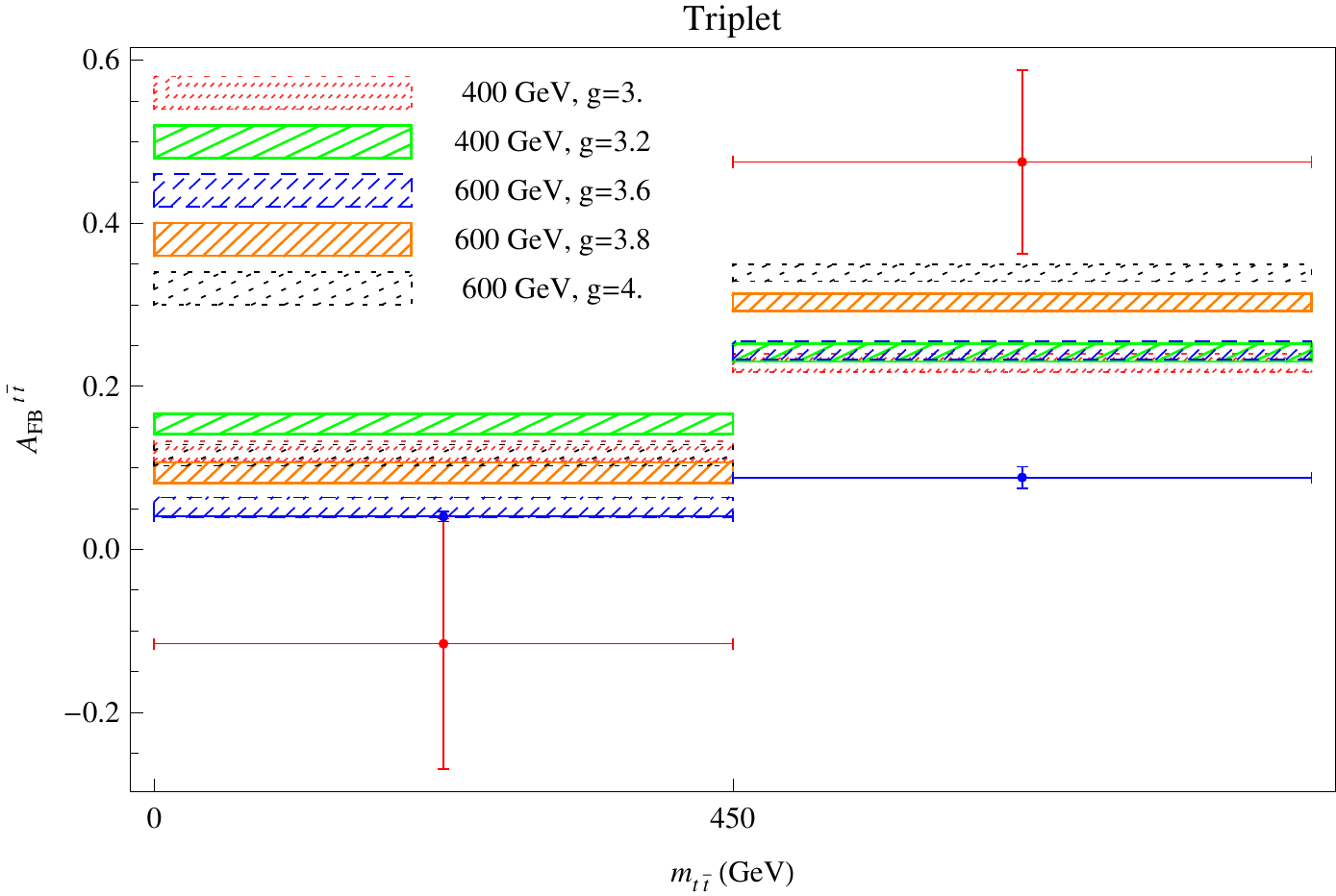}
\includegraphics[width=0.45\textwidth]{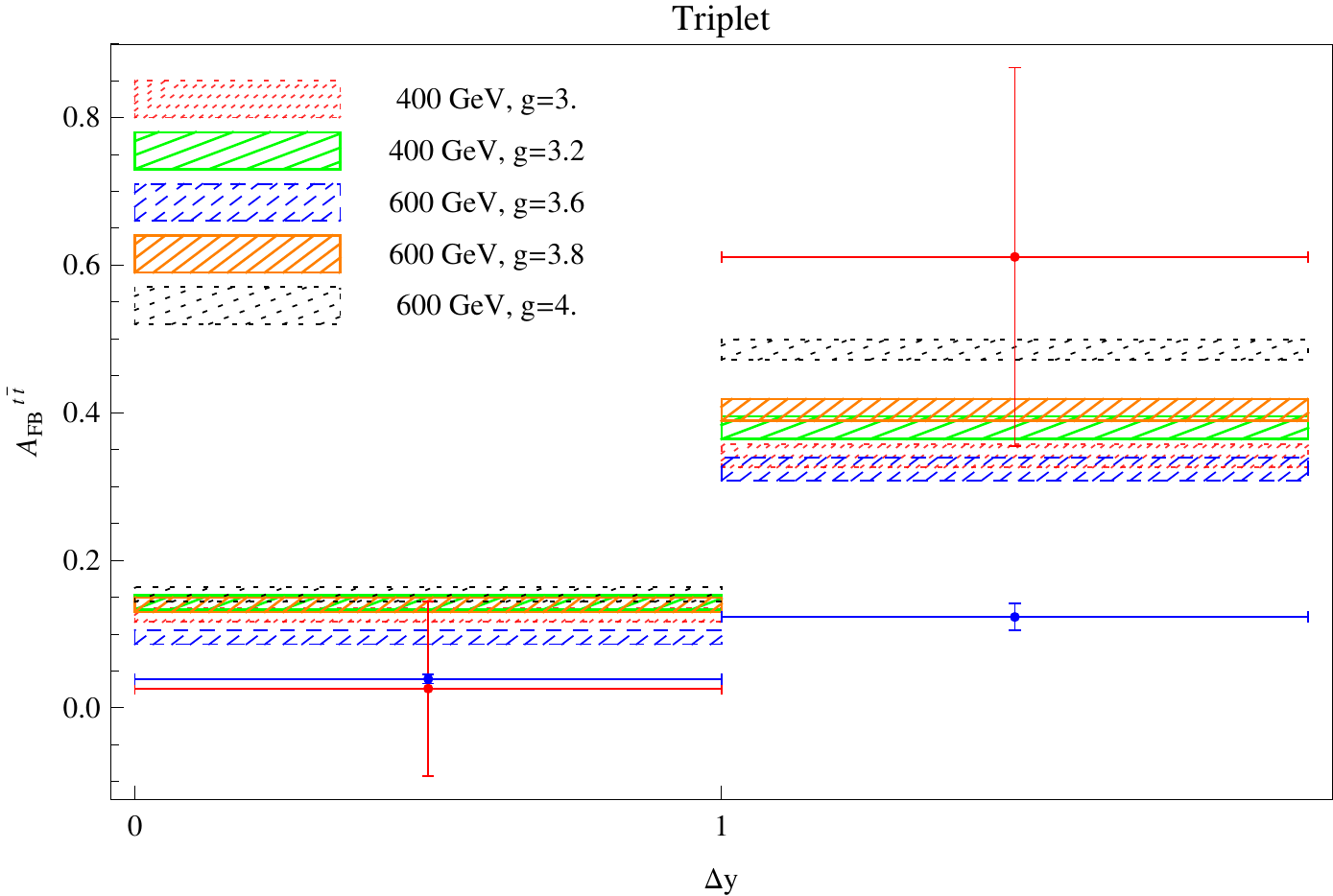}\\
\includegraphics[width=0.45\textwidth]{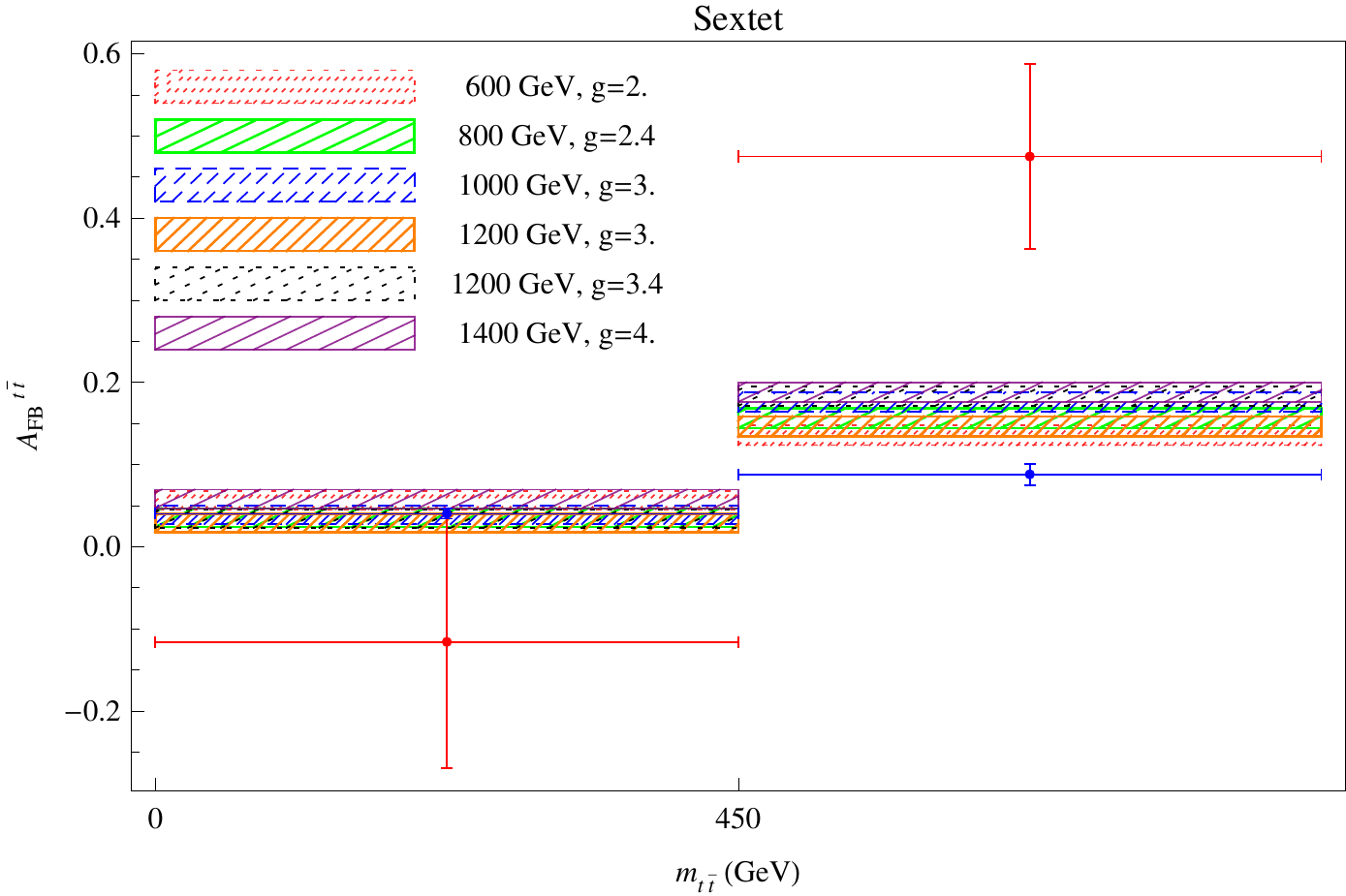}
\includegraphics[width=0.45\textwidth]{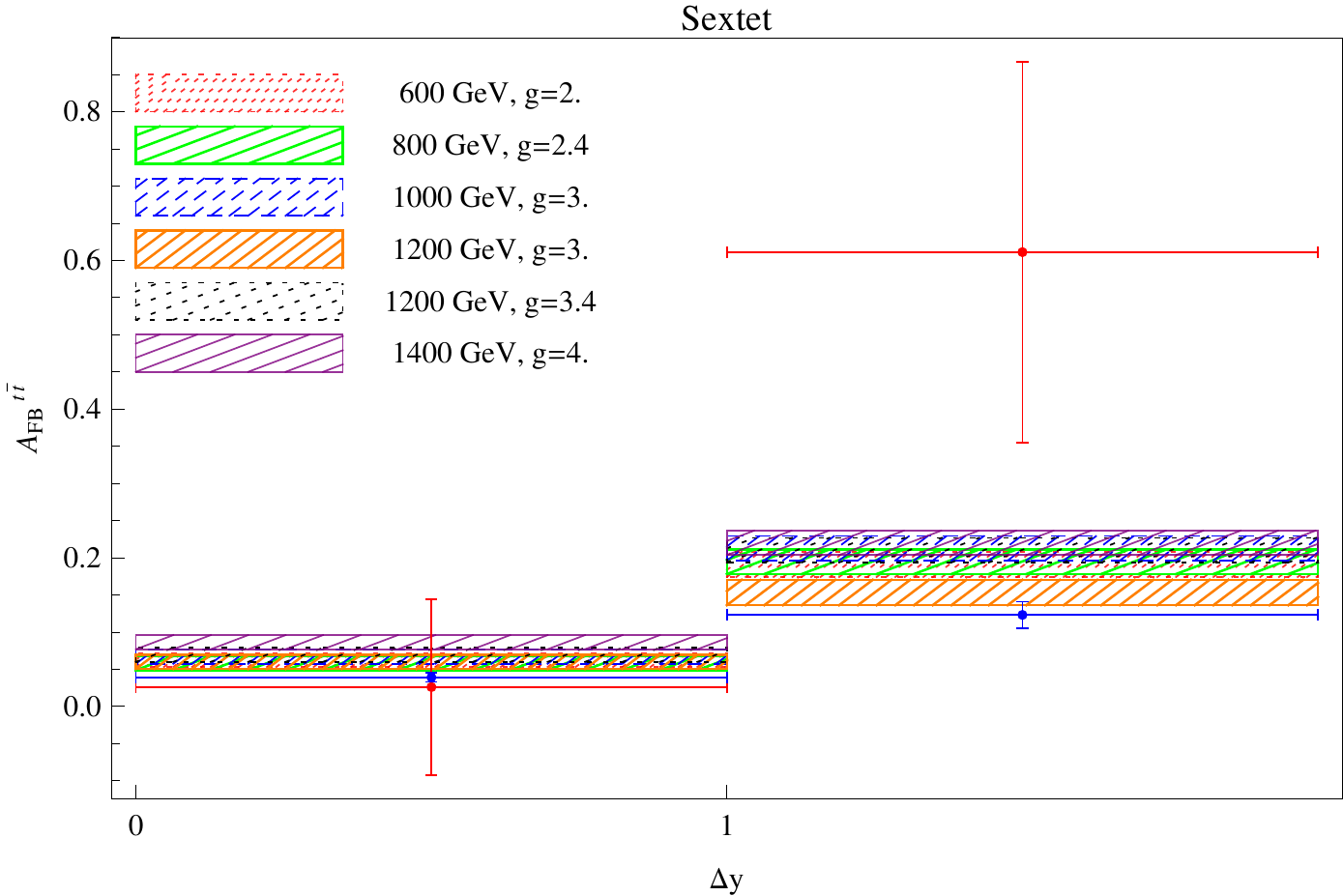}\\
\includegraphics[width=0.45\textwidth]{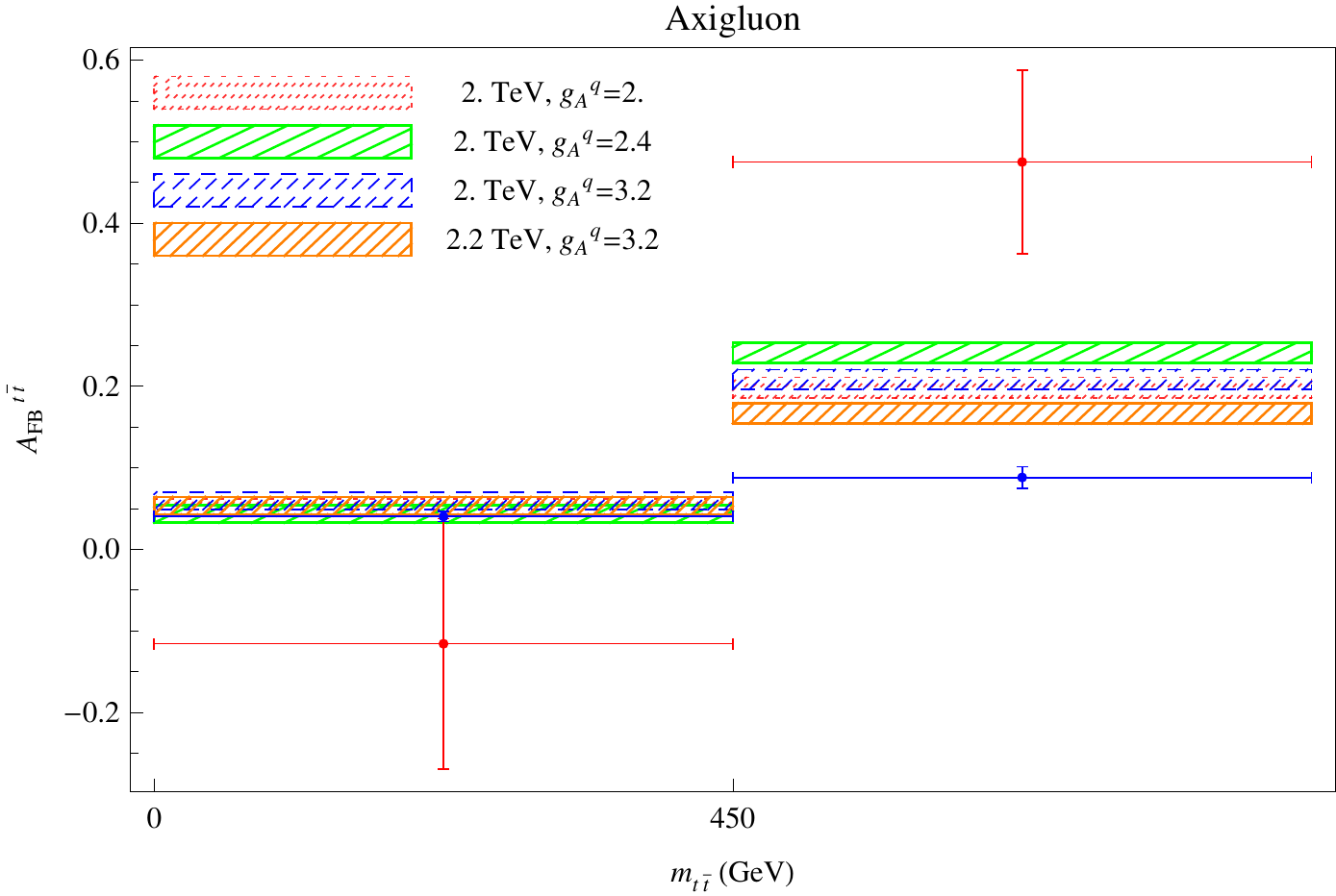}
\includegraphics[width=0.45\textwidth]{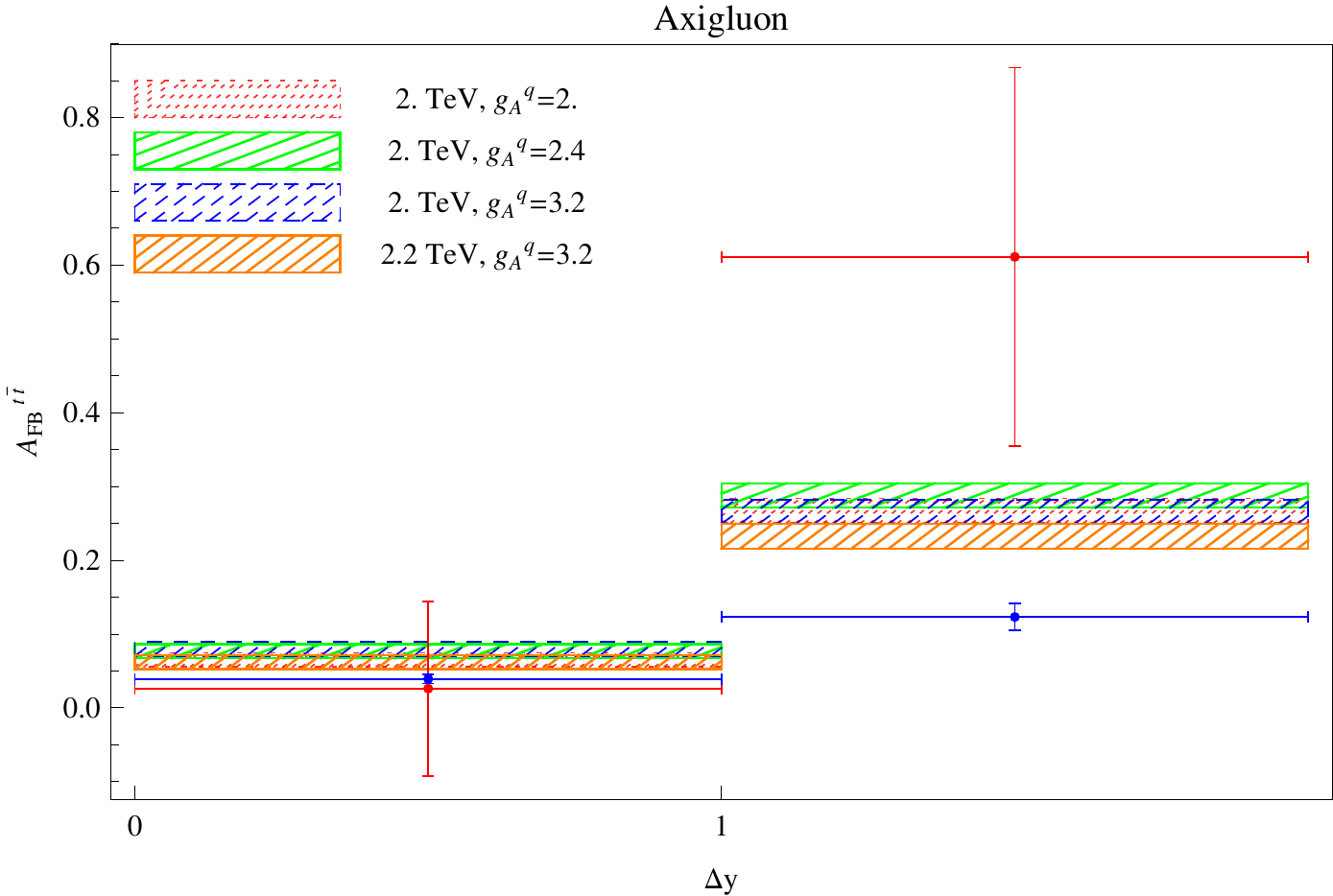}
\end{center}
\caption[Parton-level $A_{FB}$ for triplet, sextet and axigluon models.]{$A_{FB}$ for triplet, sextet and axigluon models with couplings as indicated in the legend, with the hatched regions corresponding to $1\sigma$ errors based on the limited statistics of the sample. The red bars are the CDF observation with $1\sigma$ errors, while the blue bars indicate the NLO SM contribution, which has not been included in the LO contribution calculated via {\tt MadGraph} and {\tt PYTHIA}. The last bin includes all events with $m_{t \bar{t}} > 450$ GeV and $|\Delta y| > 1$, respectively.}
\label{TripMttDely}
\end{figure}

\begin{figure}
\begin{center}
\includegraphics[width=0.45\textwidth]{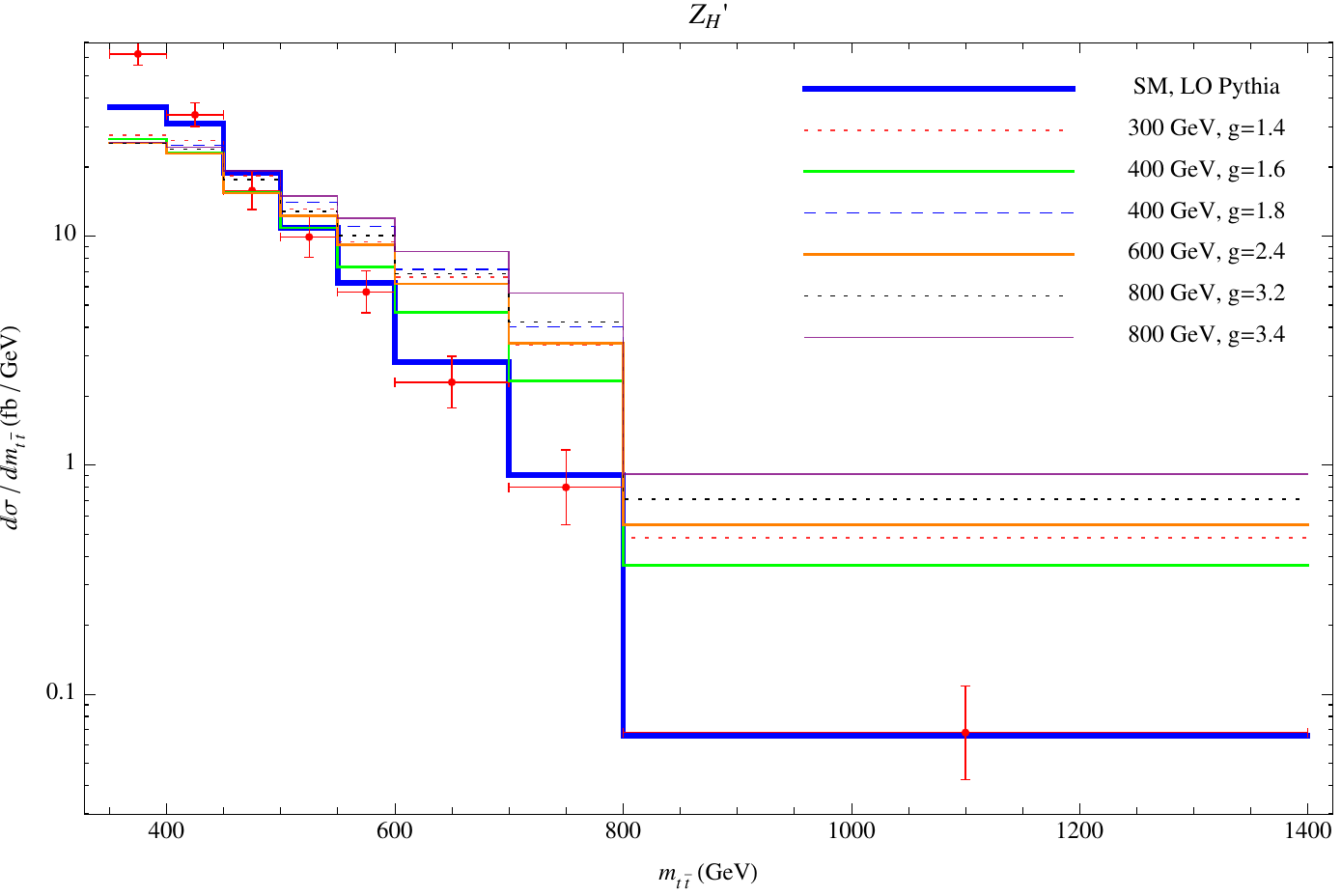}
\includegraphics[width=0.45\textwidth]{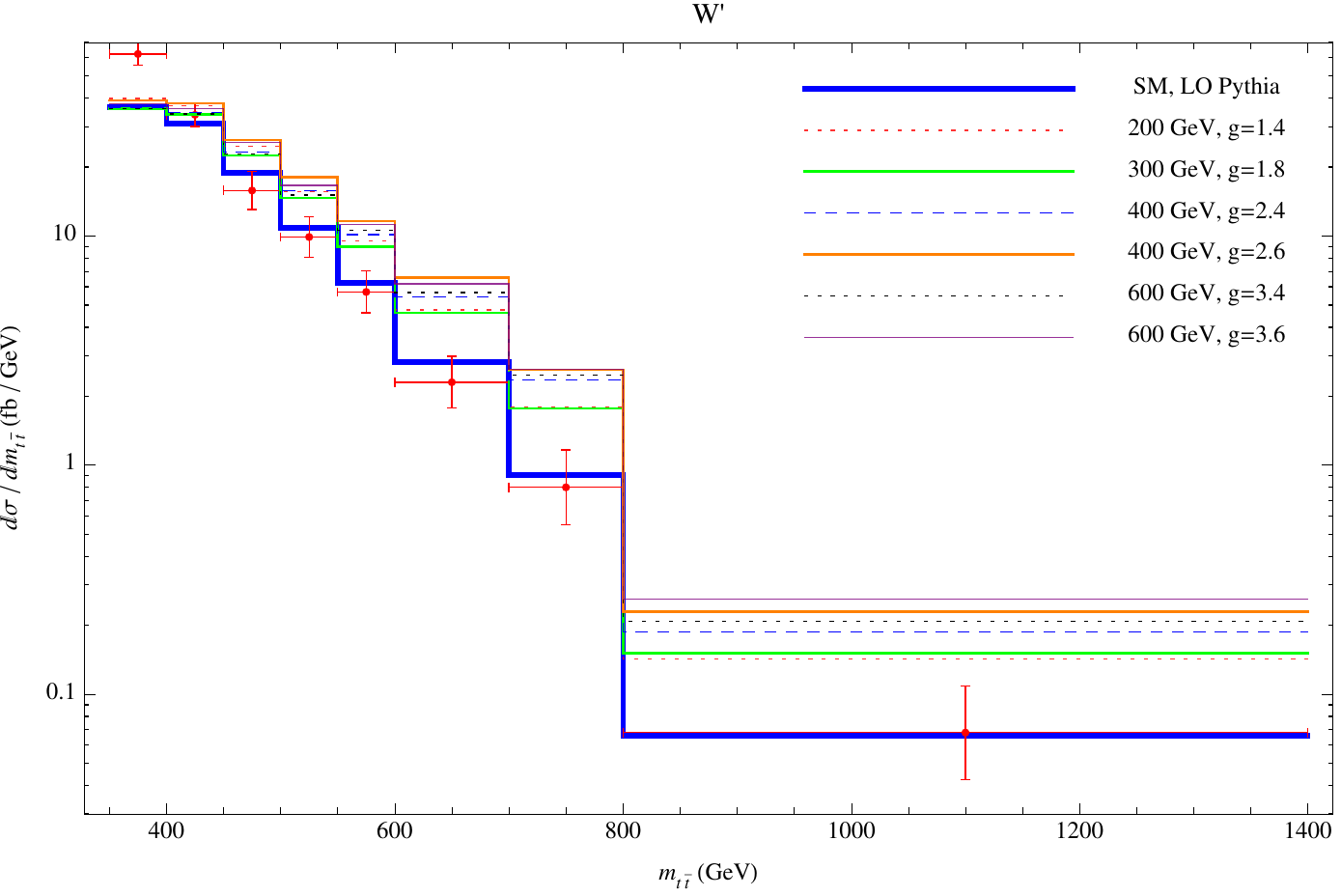}\\
\includegraphics[width=0.45\textwidth]{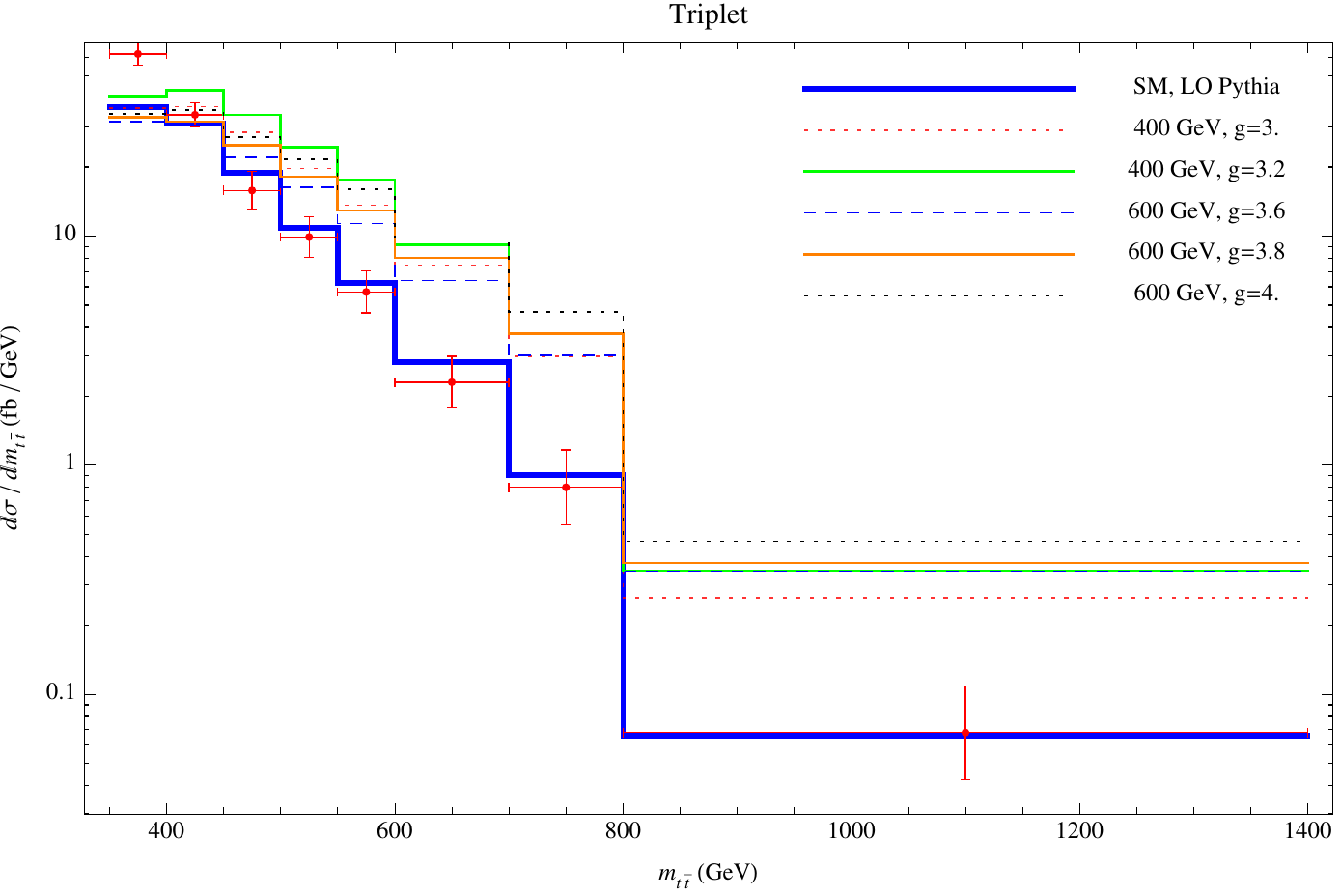}
\includegraphics[width=0.45\textwidth]{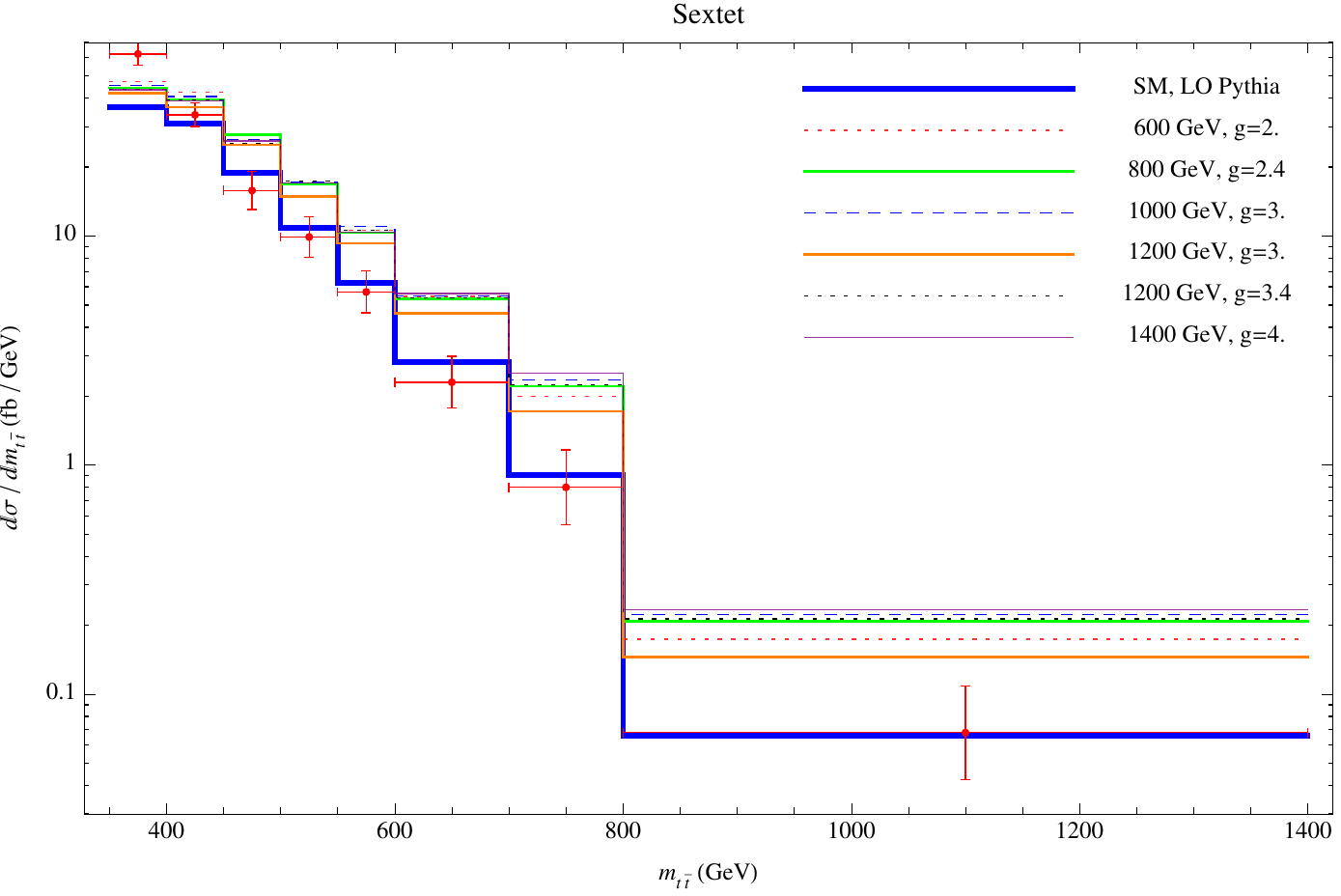}\\
\includegraphics[width=0.45\textwidth]{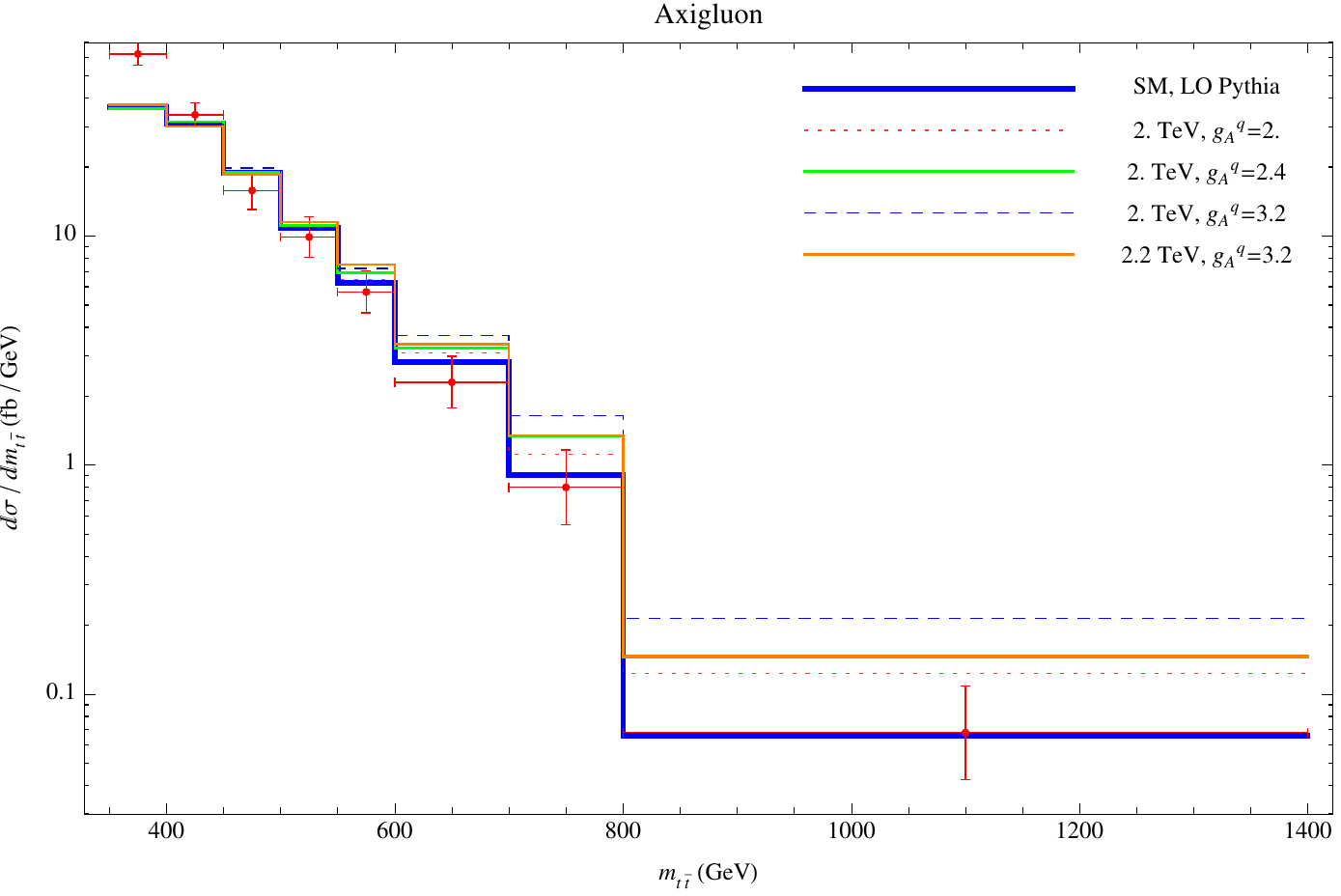}
\end{center}
\caption[Parton-level $d \sigma \over d m_{t \bar{t}}$ for the benchmark models.]{$d \sigma \over d m_{t \bar{t}}$ for the benchmark models appearing in Figs.~(\ref{ZpWpMttDely},~\ref{TripMttDely}). The Tevatron measured cross section (red crosses, from \cite{Aaltonen:2009iz}), and  LO SM cross-section with the same SM parameters and fixed renormalization scale used to generate benchmark model events are also shown. No K-factors are applied.}
\label{InvariantMassDistribution}
\end{figure}

A similar analysis is carried out for triplets, sextets and axigluons in Fig.~(\ref{TripMttDely}).  While triplet and sextet models can marginally reproduce the asymmetries at LO, the rise between the low and high invariant mass bins and low and high rapidity bins is not as pronounced for the triplets and sextets as for the $W'$ and $Z'_H$.  The reason for this in the sextet and triplet cases can, for example, be easily extracted from the analytical expressions, Eqs.~(\ref{Z'int}),~(\ref{Z'sq}),~(\ref{tripintsq}).  At high invariant mass, the scattering amplitude is dominated by the squared term.  There $\hat{u}_t \simeq \hat{u}_\phi$ in Eq.~(\ref{tripintsq}), and the effect of the ${\cal A}_{sq}$ on the asymmetry vanishes.  By contrast, the $Z'_H,~W'$ are dominated by $\hat{u}_t^2/\hat{t}_M^2 \sim (1+c_\theta)^2/(1-c_\theta)^2$ which retains a contribution to the asymmetry at high invariant mass.

The axigluon models tend to significantly underproduce the asymmetry in the high invariant mass window.  Choosing a larger coupling does not give rise to a larger asymmetry in the high invariant mass bin because of width effects, and the axigluon mass cannot be lowered in order to compensate on account of dijet constraints \cite{Bai:2011ed}.  Thus we see that axigluon models have greater difficulty than $Z'_H$ and $W'$ for reproducing the observations.  Some of these constraints can be relaxed somewhat by moving away from the point $g_A^q = - g_A^t$ \cite{Bai:2011ed}.   In addition, on account of the large couplings present in these models, NLO corrections to the new physics must be considered in order to draw firm conclusions.

Before moving on to the fully reconstructed sample, we compare the invariant mass distributions of the LO {\tt PYTHIA} results against the observations in Fig.~(\ref{InvariantMassDistribution}).  No K-factors for NLO corrections have been applied.  All the models overproduce the extracted invariant mass spectrum in the high mass bins.  Here again, however, the caveat must be applied that the $Z'_H$ and $W'$ models have lower selection efficiencies in the high invariant mass bins, so that one expects the discrepancy in the bins above $m_{t\bar{t}} \approx 500$ GeV to be greatly reduced for these models.  On the other hand, the sextets and triplets severely overproduce the observed number of events and, based on Fig.~(\ref{contours}), are not helped by having lower selection efficiency in the high invariant mass bins.   The axigluon models do not have as severe an overproduction problem, but also do not generate a large asymmetry, as can be seen in Fig.~(\ref{TripMttDely}).   As commented earlier, it is of course possible that NLO corrections from the new physics will lead to significant changes in these distributions on account of the large couplings.

We now turn to the reconstructed sample from which it will be possible to make more quantitative statements about the observed versus model-dependent predicted invariant mass spectra and asymmetries.

\section{Fully Reconstructed Asymmetry and Invariant Mass Distributions}
\label{Sec:reconstructed}

To reconstruct the invariant mass spectrum and check the model dependence of the $t\bar{t}$ parton-level asymmetry extracted in \cite{Aaltonen:2011kc} for the class of models discussed here, we can send the showered $t\bar{t}$ events through {\tt PGS}. To select the $t\bar{t}$ signal, we take the same requirements as CDF in their analysis: 
\begin{itemize}
\item Exactly one electron or muon with $p_T > 20$ GeV and $|\eta| < 1.0$.
\item Photon and $\tau$ veto.
\item At least four jets with $p_T > 20$ GeV and $|\eta| < 2.0$, with at least one of the jets having a $b$-tag.  
\item $E_T^{miss} > 20$ GeV. 
\end{itemize}
We must then fully reconstruct the decayed tops.  We do a likelihood analysis on the lepton and jet kinematics to the $t \bar{t}$ hypothesis, using the algorithm described in our previous paper \cite{Gresham:2011dg}.  The top is reconstructed out of the four hardest jets in the event.   In order to gain enough statistics in the high invariant mass bin to reliably compare our results against the reconstructed CDF asymmetry, we generate 5 million $t\bar{t}$ events per model.    Approximately 2\% of these events survive the cuts.  In contrast to our previous analysis \cite{Gresham:2011dg}, but in accordance with the CDF analysis, we place no $\chi^2$ cut on the reconstruction of the tops.  We explore later the effect of the $\chi^2$ cut on the size of the asymmetry.

Because of the large numbers of simulated and reconstructed events required, we consider only a representative subset of the models analyzed at the parton level in the previous section.  We choose a 400 GeV $Z'_H$ with $g_R = 1.75$, a 400 GeV $W'$ with $g_R = 2.55$, a 600 GeV triplet with $g = 4.0$, a 1.4 TeV sextet with $g = 4.0$ and a 2.0 TeV axigluon with $g_A^q = -g_A^t = 2.4$.\footnote{Note that the $W'$ and $Z'_H$ models will require a triplet or higher Higgs representation in order to evade dijet constraints for the flavor-conserving $Z'$s, which exist in these models.  On the other hand, we also compare a heavier 800 GeV $Z'_H$ to the lighter $W'$ and $Z'_H$, and find that it overproduces the high invariant mass spectrum.}  As we will see, the CDF extraction of the parton level asymmetries and invariant mass spectra is somewhat model dependent, on account of the model dependent efficiencies shown in Fig.~(\ref{efficiencies}) and Table~(\ref{coarse bin efficiencies}), as well as detector effects.

\begin{figure}
\begin{center}
\includegraphics[width=0.48\textwidth]{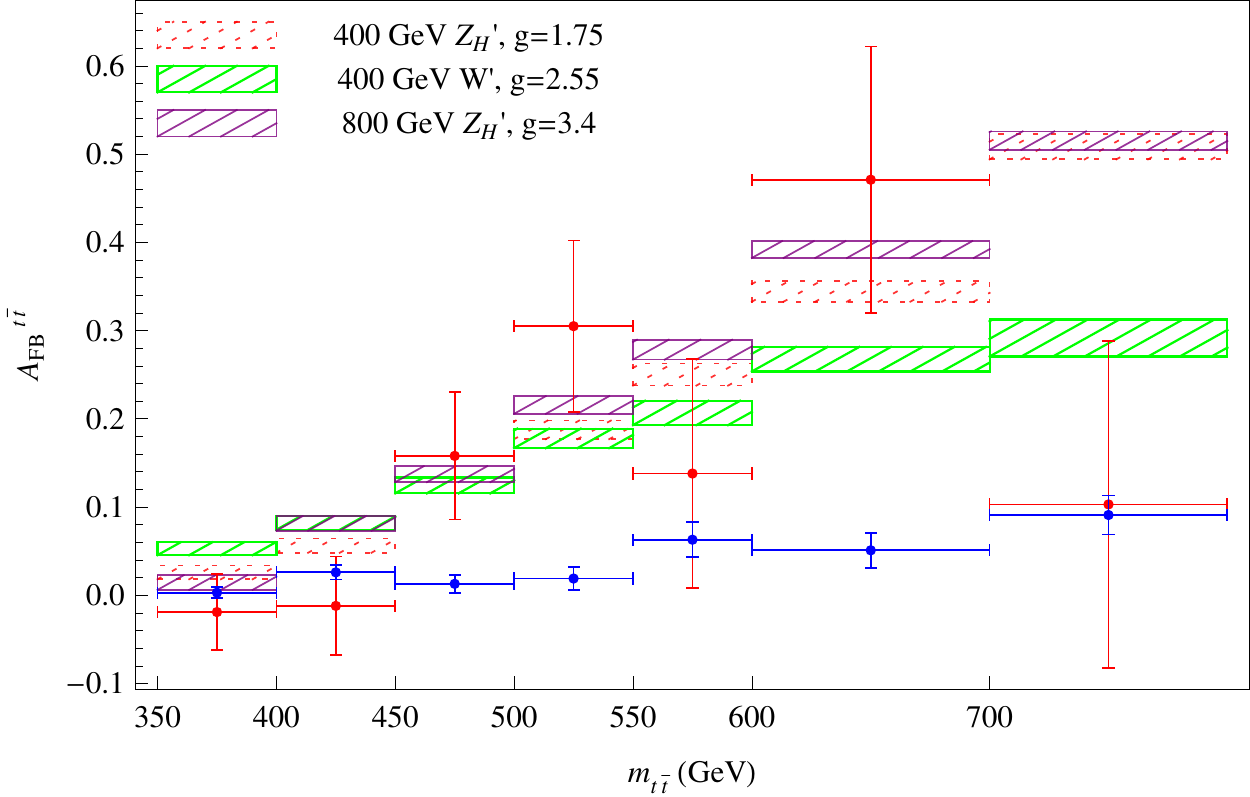}
\includegraphics[width=0.48\textwidth]{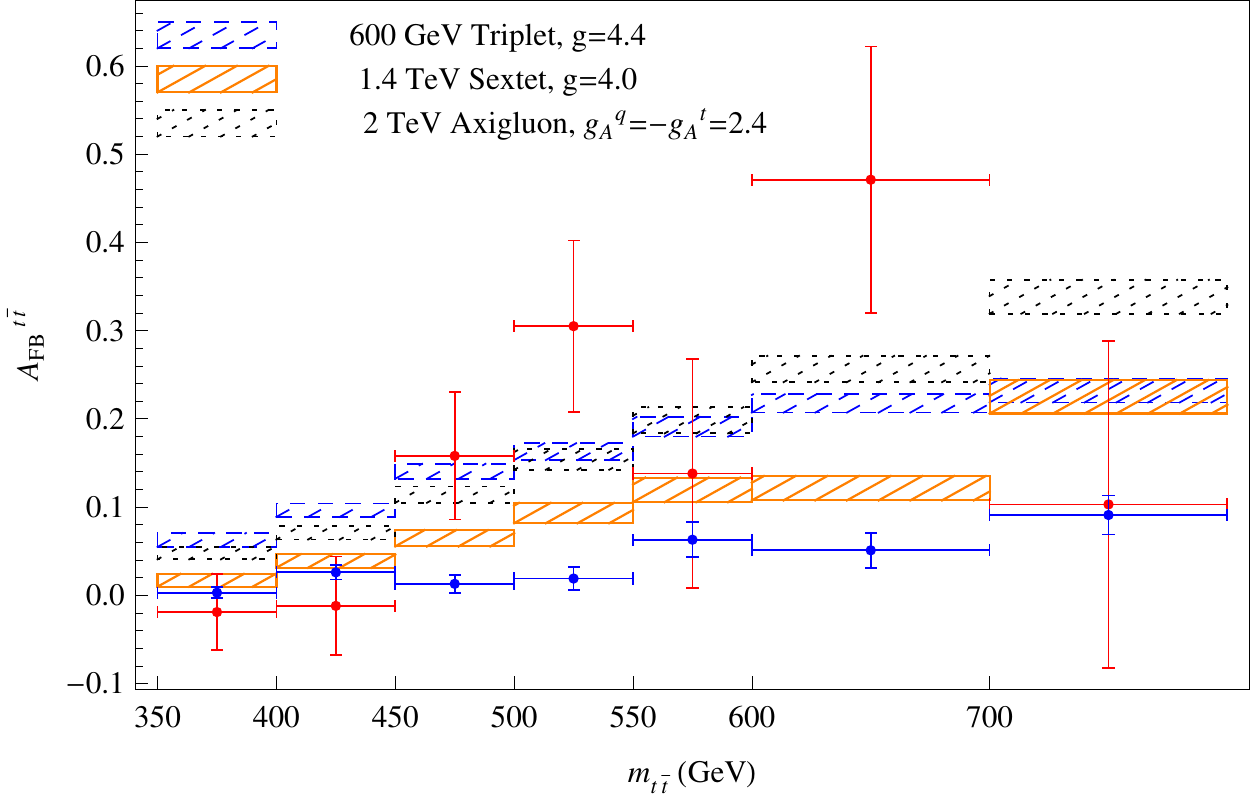}\end{center}
\caption[Reconstructed $A_{FB}^{t \bar{t}}(M_{t \bar{t},i})$ for benchmark models.]{$A_{FB}^{t \bar{t}}(M_{t \bar{t},i})$ for $W'$, $Z'_H$, triplet, sextet and axigluon models. Red crosses are the CDF values reconstructed from data. Blue crosses are the MC@NLO expectation. The last bin includes all events with $m_{t \bar{t}} > 700$ GeV.}
\label{ReconstructedWpZp}
\end{figure}

\begin{figure}
\begin{center}
\includegraphics[width=0.45\textwidth]{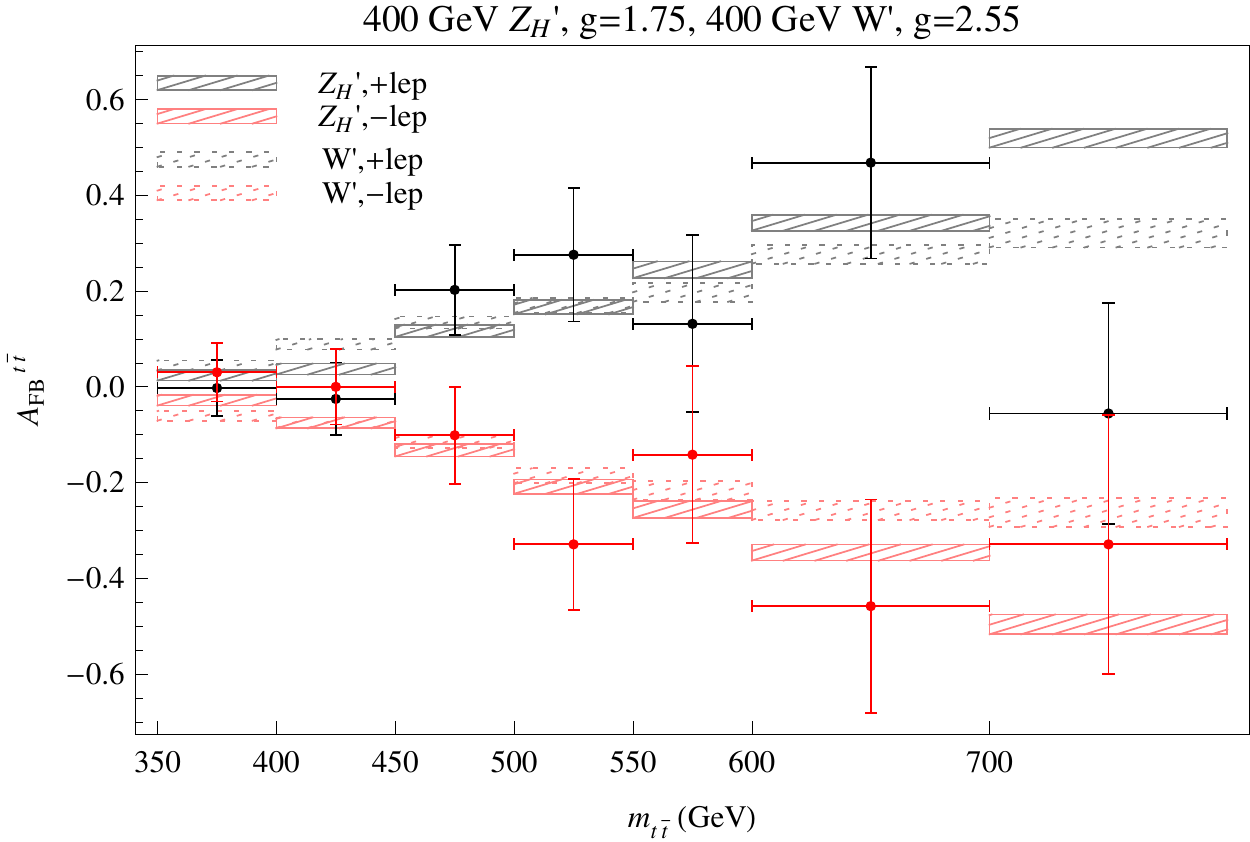}
\includegraphics[width=0.45\textwidth]{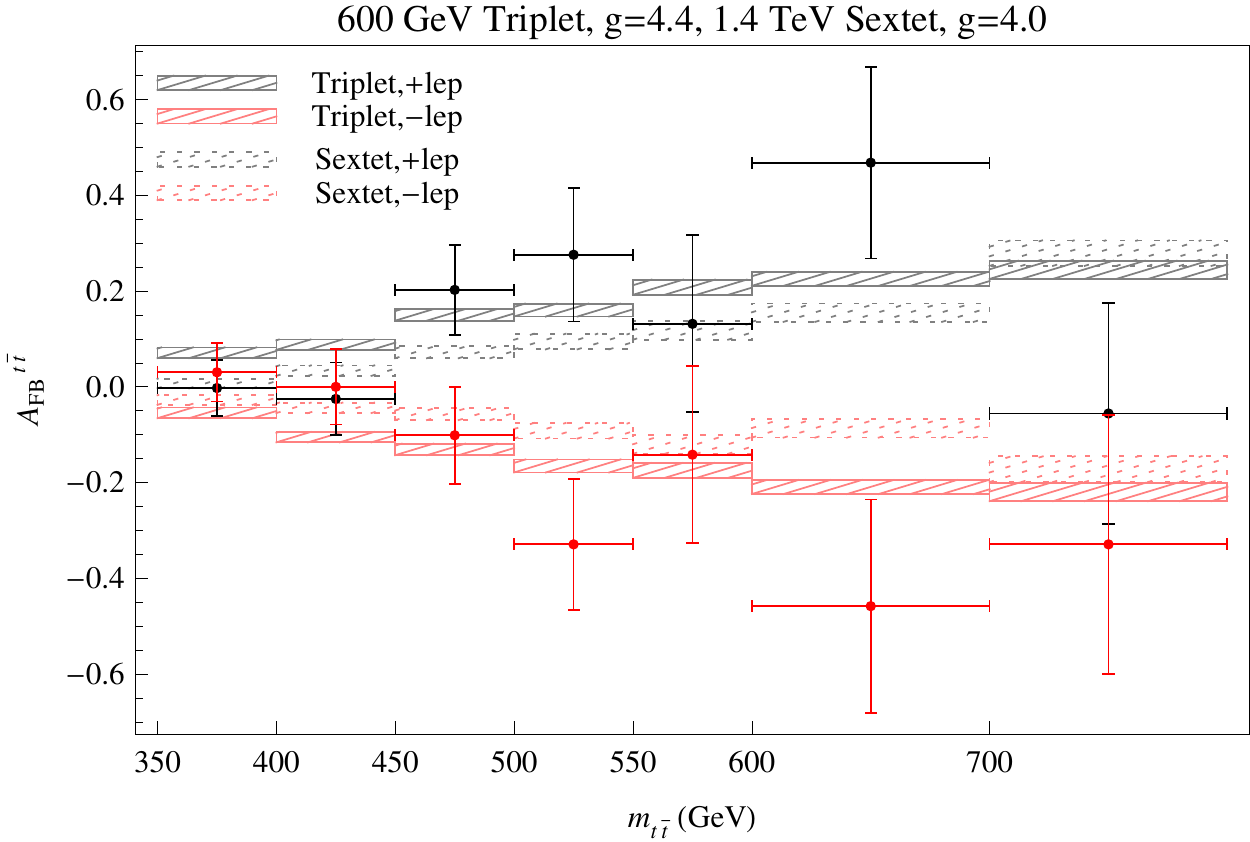} \\
\includegraphics[width=0.45\textwidth]{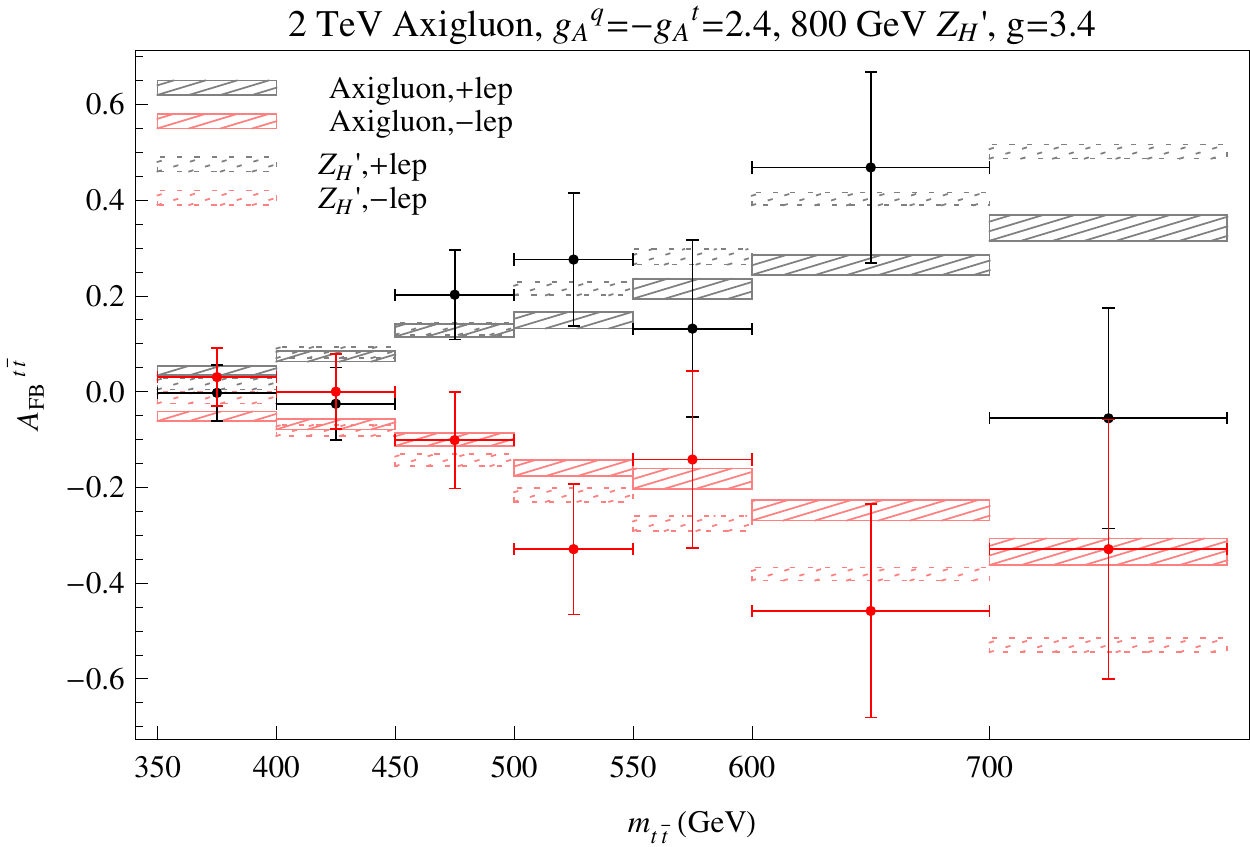}
\end{center}
\caption[Reconstructed $ A_{FB}^{t \bar{t}}(q, M_{t \bar{t}})$ for benchmark models.]{$ A_{FB}^{t \bar{t}}(q, M_{t \bar{t}})$ as defined in Eq.\eqref{AFBdeflep} for $W'$, $Z_H'$, triplet, sextet, and axigluon models. Here, the data is divided according to the charge, $q$, of the lepton in the event. Black ($q >0$) and red ($q < 0$) crosses are the CDF values reconstructed from data (with background subtracted).}
\label{ReconstructedWpZpTripSixPartitioned}
\end{figure}

We begin by comparing our reconstructed results against the CDF results for several models in Fig.~(\ref{ReconstructedWpZp}).  The $A_{FB}^{t\bar{t}}$ shown there is as defined in Eq.~(\ref{AFBdef}), with the top and anti-top identified by the sign of the lepton.  We see that models that reproduce the parton level asymmetry match well against the fully reconstructed CDF asymmetry.  The $Z'_H$ and $W'$ models, however, receive a larger upward correction upon unfolding to the parton level than would be expected using SM efficiencies.  The reason for this is clear from Table~(\ref{coarse bin efficiencies}): the efficiencies in the high invariant mass bin with $\Delta y > 0$ for $Z'_H$ and $W'$ models are lower than for the SM, leading to a greater washout of the asymmetry at the detector level.  We also observe that the axigluon, while appearing to reproduce the reconstructed asymmetry marginally, tends to underproduce the unfolded parton level asymmetry as seen in Fig.~(\ref{TripMttDely}).  We can also look at the partitioned asymmetry defined by
\begin{equation}
A^{t\bar{t}}(q, M_{t\bar{t},i}) = \frac{N((y_\ell - y_h)>0, q, M_{t\bar{t},i})-N((y_\ell - y_h) <0, q, M_{t\bar{t},i})}{N((y_\ell - y_h)>0, q, M_{t\bar{t},i})+N((y_\ell - y_h) <0, q, M_{t\bar{t},i})},
\label{AFBdeflep}
\end{equation}  
where $y_\ell$ is the rapidity of the leptonic top, $y_h$ is the rapidity of the hadronic top, and $q$ is the charge of the lepton.  The asymmetries obtained in this way are shown in Fig.~(\ref{ReconstructedWpZpTripSixPartitioned}). 

\begin{figure}
\begin{center}
\includegraphics[width=0.8\textwidth]{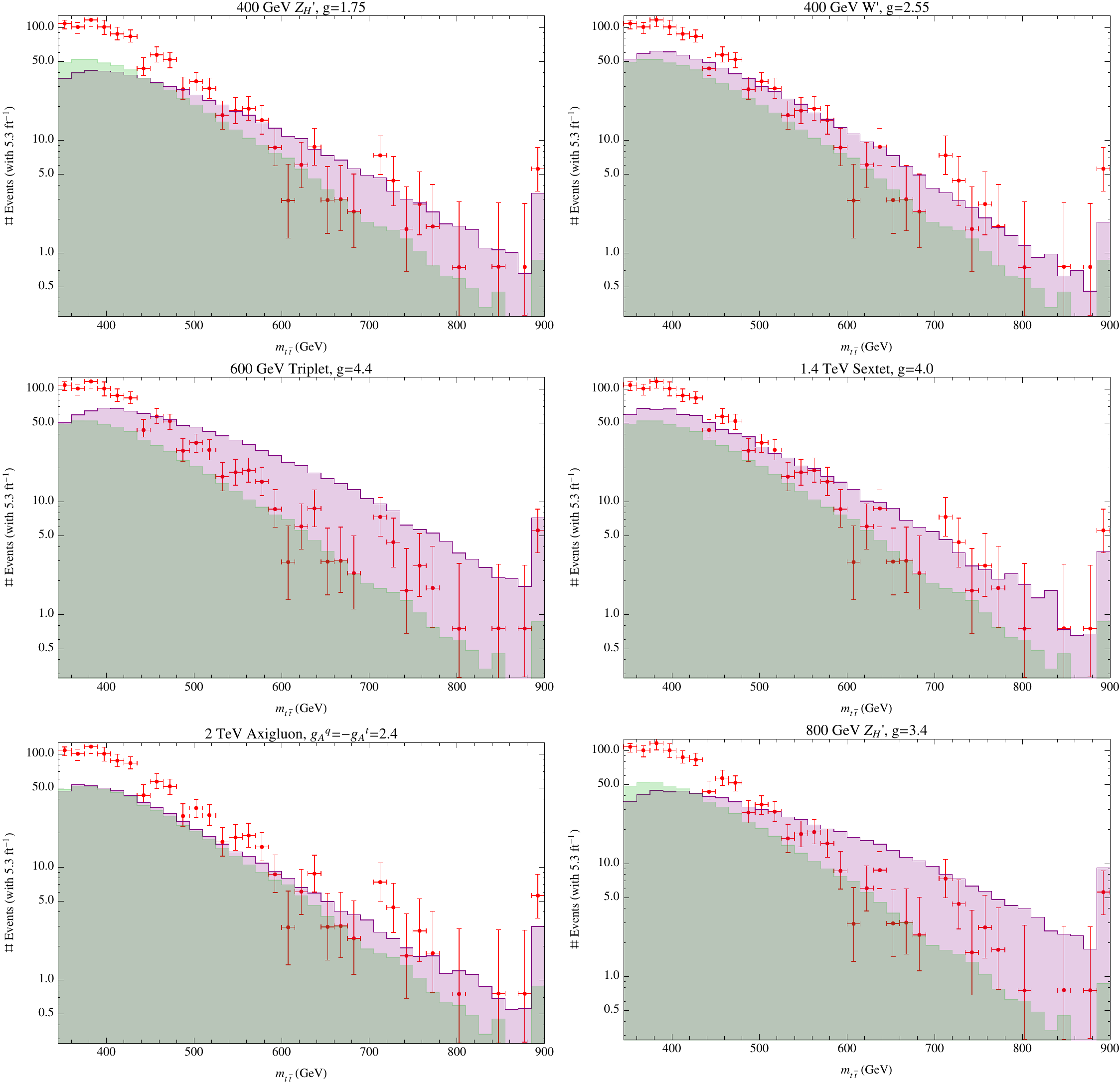}
\end{center}
\caption[Number of expected reconstructed events versus $m_{t \bar{t}}$.]{Number of expected $t \bar{t}$ events with $5.3$ fb$^{-1}$ at the Tevatron, distributed over $m_{t \bar{t}}$.  Events were passed through {\tt PGS} and then tops were reconstructed using the algorithm detailed in \cite {Gresham:2011dg}. The red bars indicate CDF's measurement, with expected background (as estimated by CDF) subtracted.  The green histogram is our SM sample and the purple histograms represent model samples. For SM and model samples, a fixed number of events were generated, and then event counts were scaled appropriately for $5.3$ fb$^{-1}$ integrated luminosity.}
\label{ReconstructedEventCounts}
\end{figure}

We also compare the reconstructed invariant mass spectrum against that reported in \cite{Aaltonen:2011kc}.  Since the efficiencies are lower for the $Z'_H$ and $W'$ than for the Standard Model, we may expect these models to agree better with the observations than suggested by the parton level invariant mass spectra shown in Fig.~(\ref{InvariantMassDistribution}).  
We compare in Fig.~(\ref{ReconstructedEventCounts}) the simulated reconstructed invariant mass spectrum against that reported in \cite{Aaltonen:2011kc}.   First we note the discrepancy at low invariant mass between all models (including the SM) and the observations, which we attribute to NLO corrections and to a difference between the {\tt PGS} detector simulation and the CDF simulation.  However, we can see the effects of the efficiencies in the high invariant mass bins noted in Fig.~(\ref{efficiencies}).  For example, we can see the efficiency correction does seem to bring the $W'$ model into agreement with the SM.  On the other hand, the triplet model largely and almost uniformly overproduces the invariant mass distribution in all bins.  While the sextet model appears in better agreement at high invariant mass, it underproduces the observed asymmetry as shown in Fig.~(\ref{ReconstructedWpZp}).  In general, triplet and sextet models have greater difficulty producing the observed asymmetry while remaining consistent with the total cross-section and invariant mass distribution, as emphasized by Figs.~(\ref{modelScatterPlot},~\ref{efficiencies}).

\begin{figure}
\begin{center}
\includegraphics[width=0.30\textwidth]{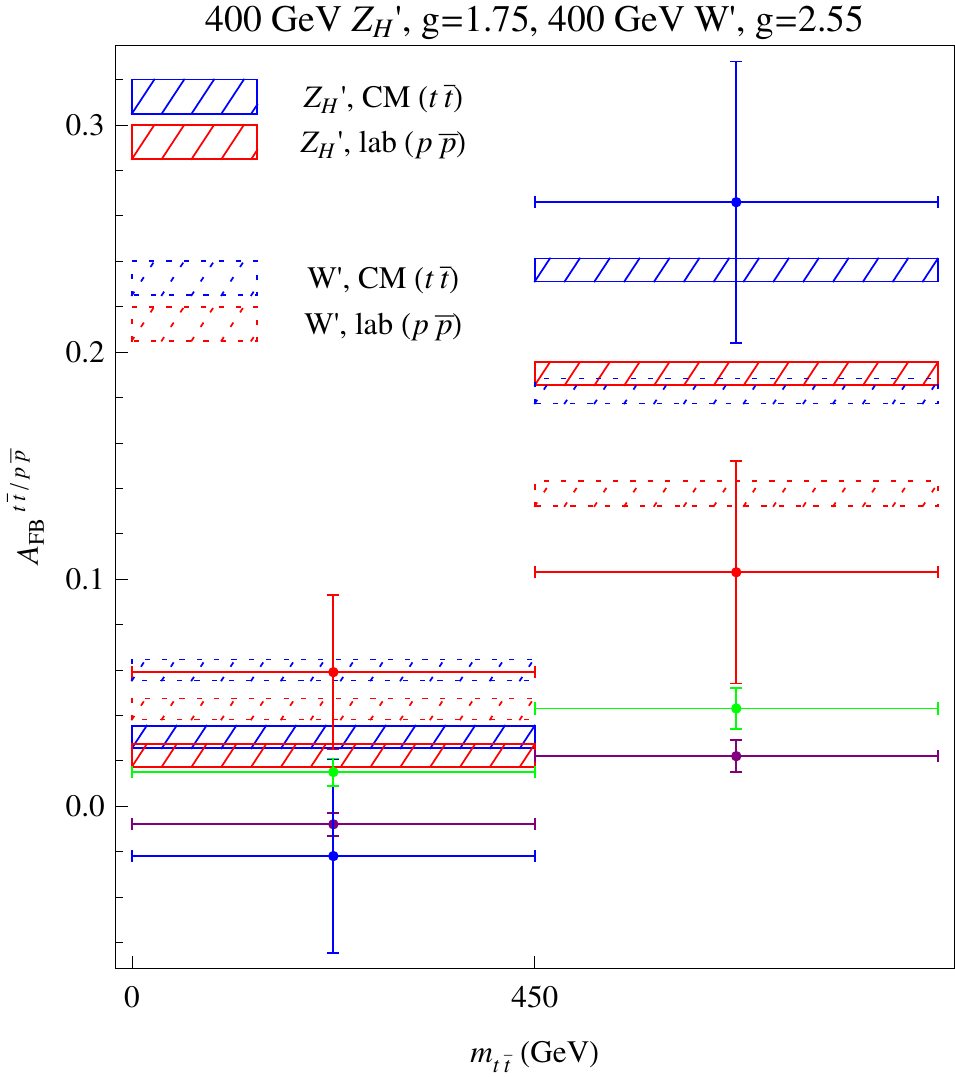}
\includegraphics[width=0.30\textwidth]{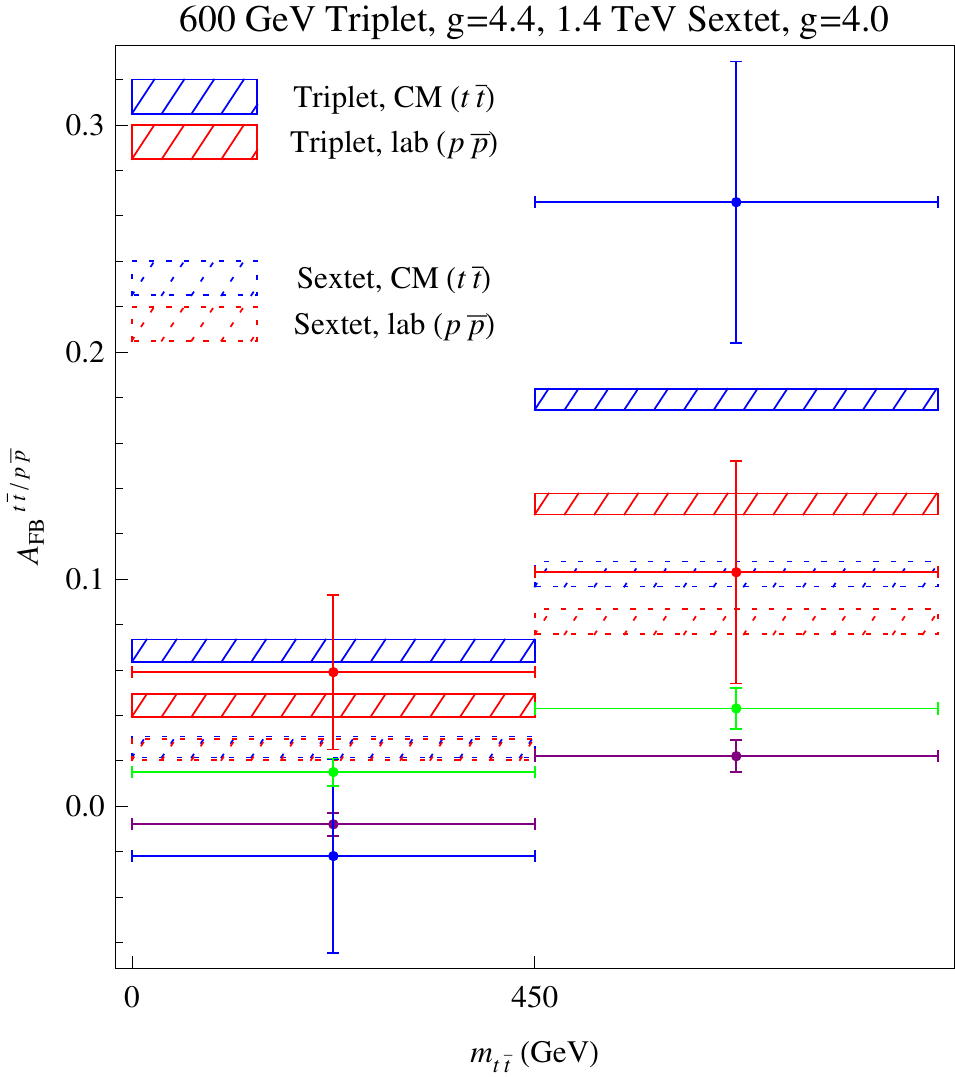}
\includegraphics[width=0.30\textwidth]{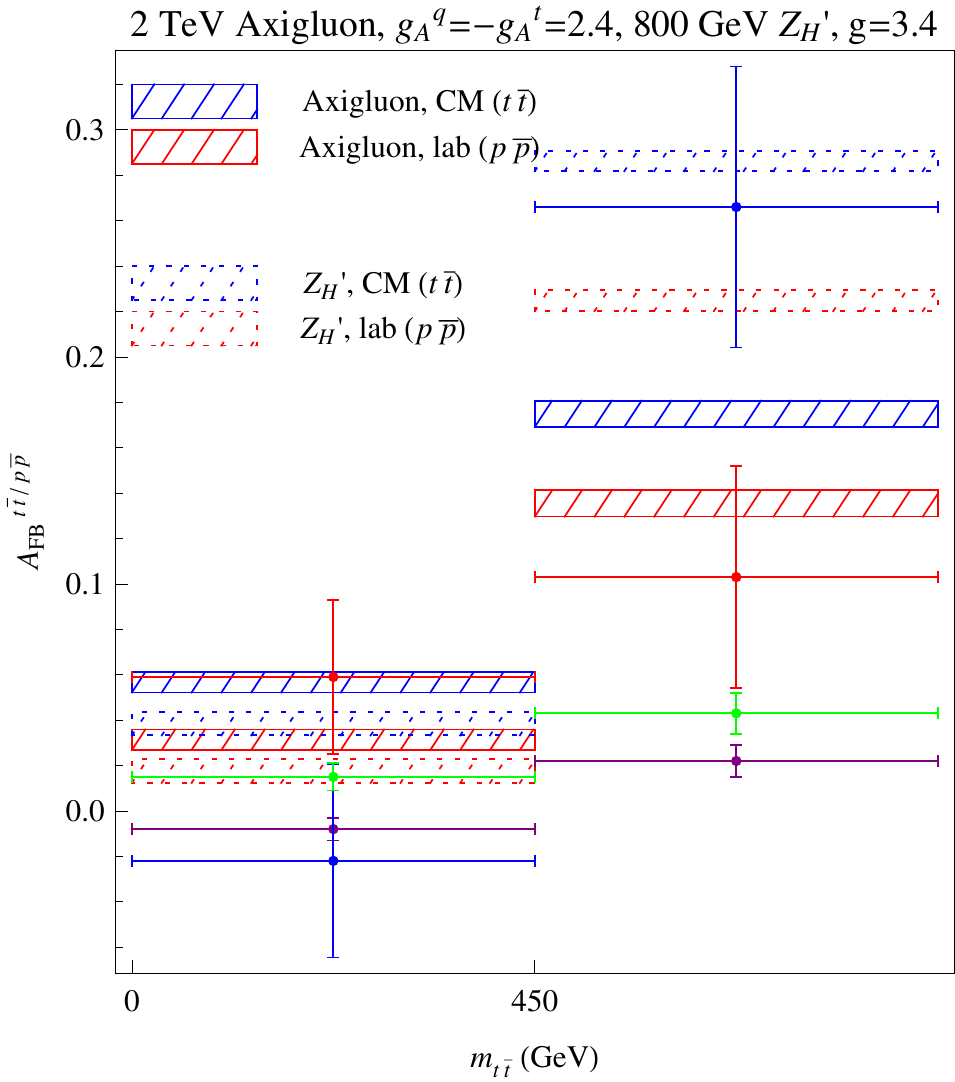}\end{center}
\caption[Reconstructed CM and lab frame $A_{FB}^{t \bar{t}}$.]{$A_{FB}^{t \bar{t}}$ and $A_{FB}^{p \bar{p}}$ in low ($m_{t \bar{t}} < 450$ GeV)  and high  ($m_{t \bar{t}} \geq 450$ GeV) $t \bar{t}$ invariant mass bins for $Z_H'$, $W'$, triplet, sextet, and axigluon models. Red / Blue crosses are the CDF values reconstructed from data in the lab / CM frames.  Purple / Green crosses indicate the SM NLO predictions.}
\label{ReconstructedWpZpTripSixAxiLabCMafb}
\end{figure}

\begin{figure}
\begin{center}
\includegraphics[width=0.30\textwidth]{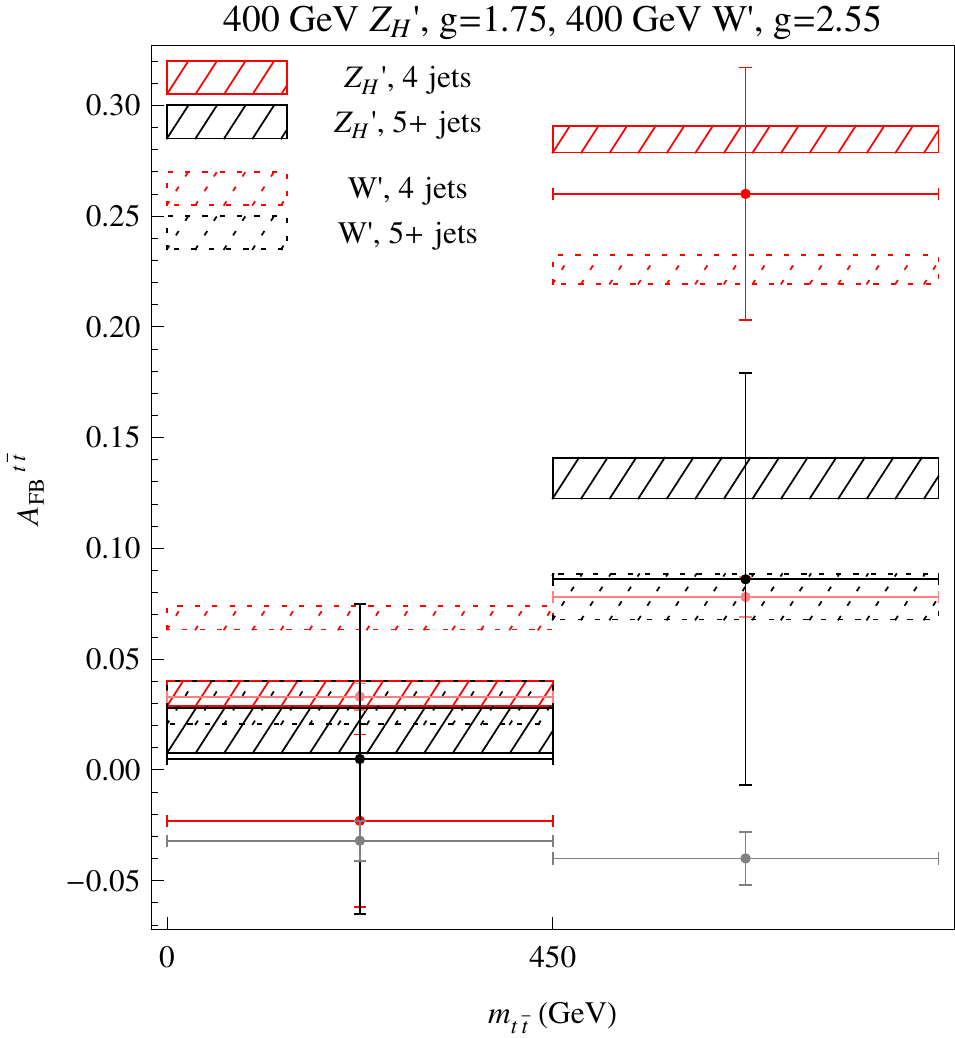}
\includegraphics[width=0.30\textwidth]{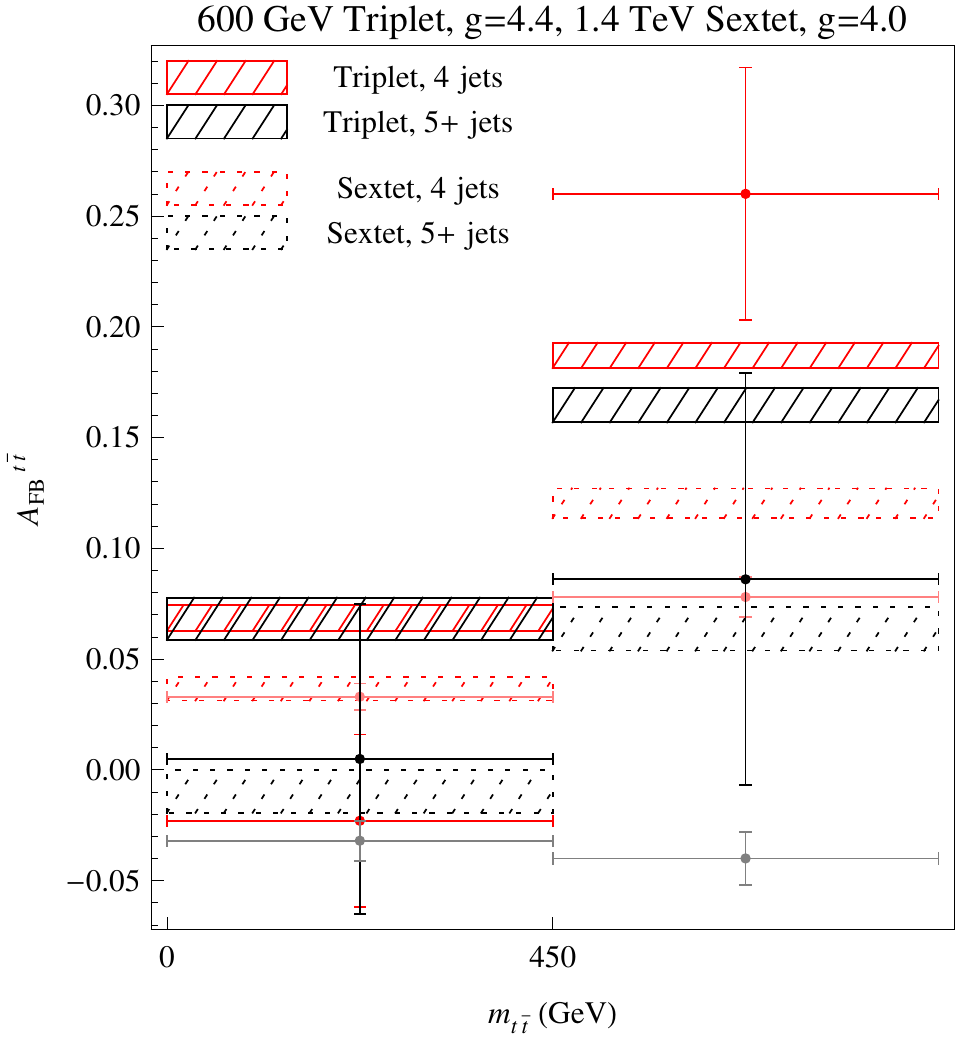}
\includegraphics[width=0.28\textwidth]{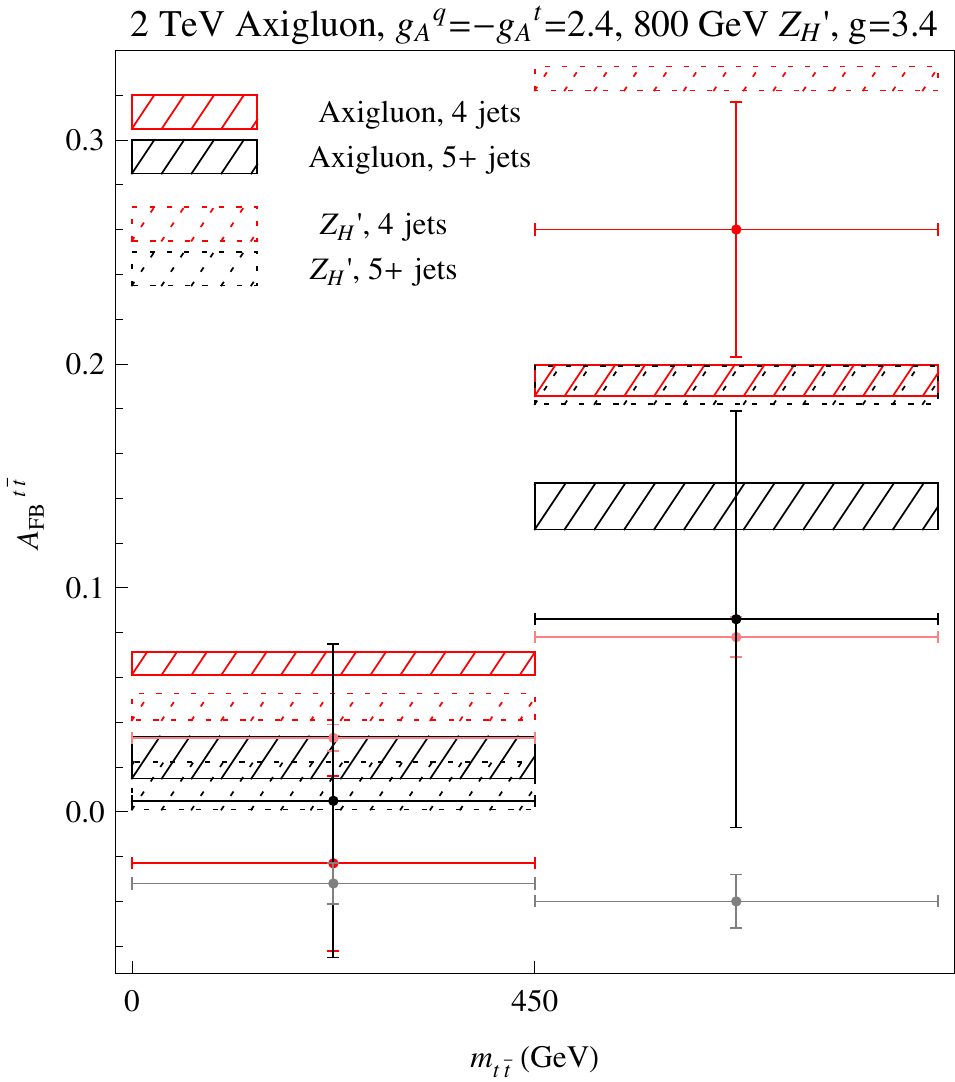}
\end{center}
\caption[Reconstructed 2-bin $ A_{FB}^{t \bar{t}}(m_{t \bar{t}})$ for 4- and 5-jet samples.]{$ A_{FB}^{t \bar{t}}$ in $t \bar{t}$ low ($m_{t \bar{t}} < 450$ GeV) and high ($m_{t \bar{t}} \geq 450$ GeV)  invariant mass bins for benchmark models. The simulated data samples were partitioned according to whether the event had more than five jets with $p_T > 20$ GeV and $|\eta| < 2$. The CDF measured lab frame asymmetry for 4-jet and 5+-jet samples is shown as red crosses and black crosses, respectively. }
\label{ReconstructedWpZpTripSixJetMult}
\end{figure}

There are two other comparisons that we are able to do with our fully reconstructed asymmetry.  We are able to compare the center-of-mass versus lab frame asymmetries, which is shown in Fig.~(\ref{ReconstructedWpZpTripSixAxiLabCMafb}).  While the models give rise to some difference between the CM and lab frames, the difference is less pronounced than what CDF observes.  
We can also compare the asymmetries in the four and five jet samples, as shown in Fig.~(\ref{ReconstructedWpZpTripSixJetMult}).  Here we see some washout of the asymmetry in the 5 jet sample, an effect that is observed in the CDF data. 

\begin{figure}
\begin{center}
\includegraphics[width=0.45\textwidth]{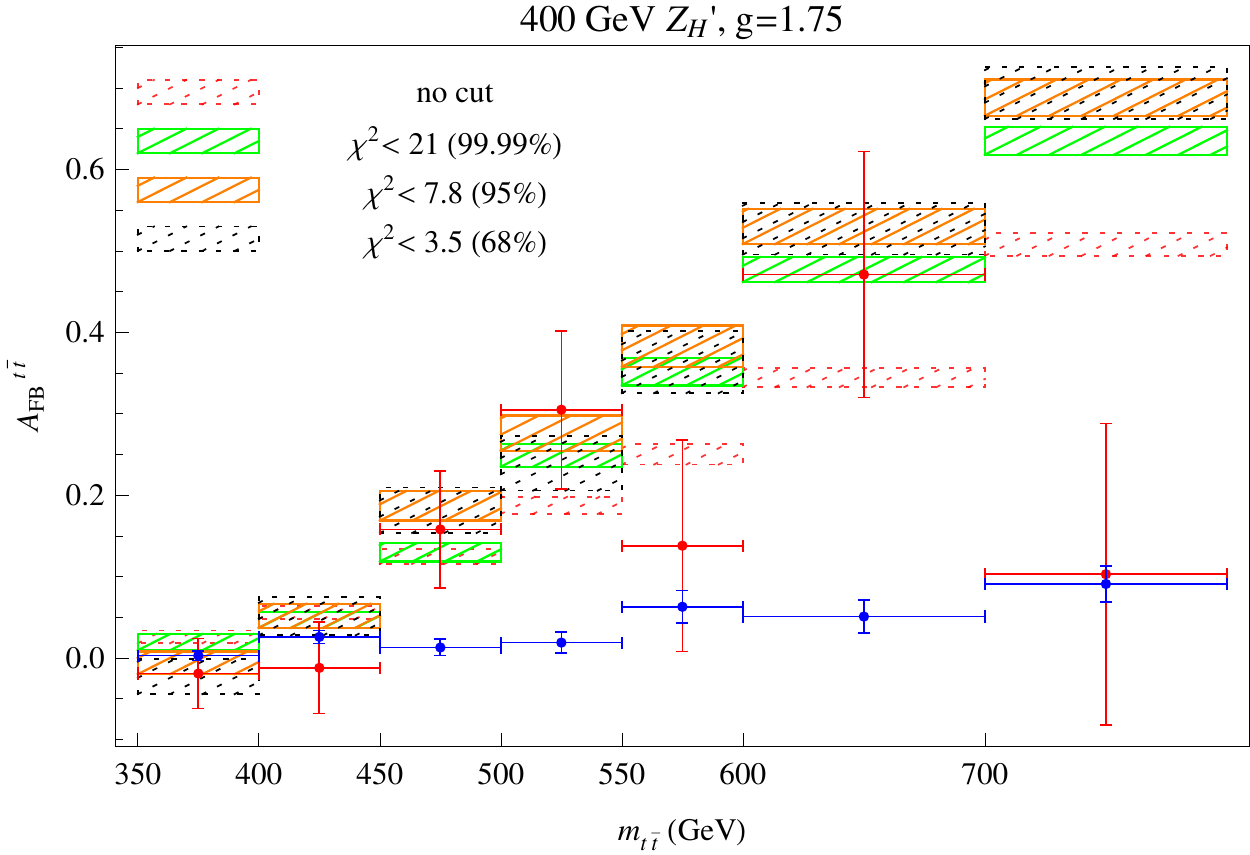}\end{center}
\caption[$A_{FB}^{t \bar{t}}(M_{t \bar{t},i})$ with various cuts on the $t \bar{t}$ reconstruction $\chi^2$.]{$A_{FB}^{t \bar{t}}(M_{t \bar{t},i})$ for $Z_H'$ model with various cuts on the $t \bar{t}$ reconstruction $\chi^2$. The last bin includes all events with $m_{t \bar{t}} > 700$ GeV. Red crosses are the CDF values reconstructed from data. CDF used a likelihood algorithm for top reconstruction, but made no $\chi^2$ cut.}
\label{ReconstructedZpTripChiSquaredCuts}
\end{figure}

Lastly, though this is not considered in detail in the CDF analysis (a value for the raw asymmetry after a $\chi^2$ cut of $3$ is presented in Table XIV of  \cite{Aaltonen:2011kc}), it is interesting to observe the effect of a $\chi^2$ cut in the top reconstruction on the size of the asymmetry.\footnote{Recall that we reconstruct tops by doing a $\chi^2$ fit on the lepton and jet kinematics to the $t \bar{t}$ hypothesis. The fit has three degrees of freedom.}  We can see in Fig.~(\ref{ReconstructedZpTripChiSquaredCuts}) that a moderate $\chi^2$ cut increases the asymmetry especially in high invariant mass bins.

\section{Lepton Asymmetry}

\begin{figure}
\begin{center}
\subfigure[]{\includegraphics[width=0.45\textwidth]{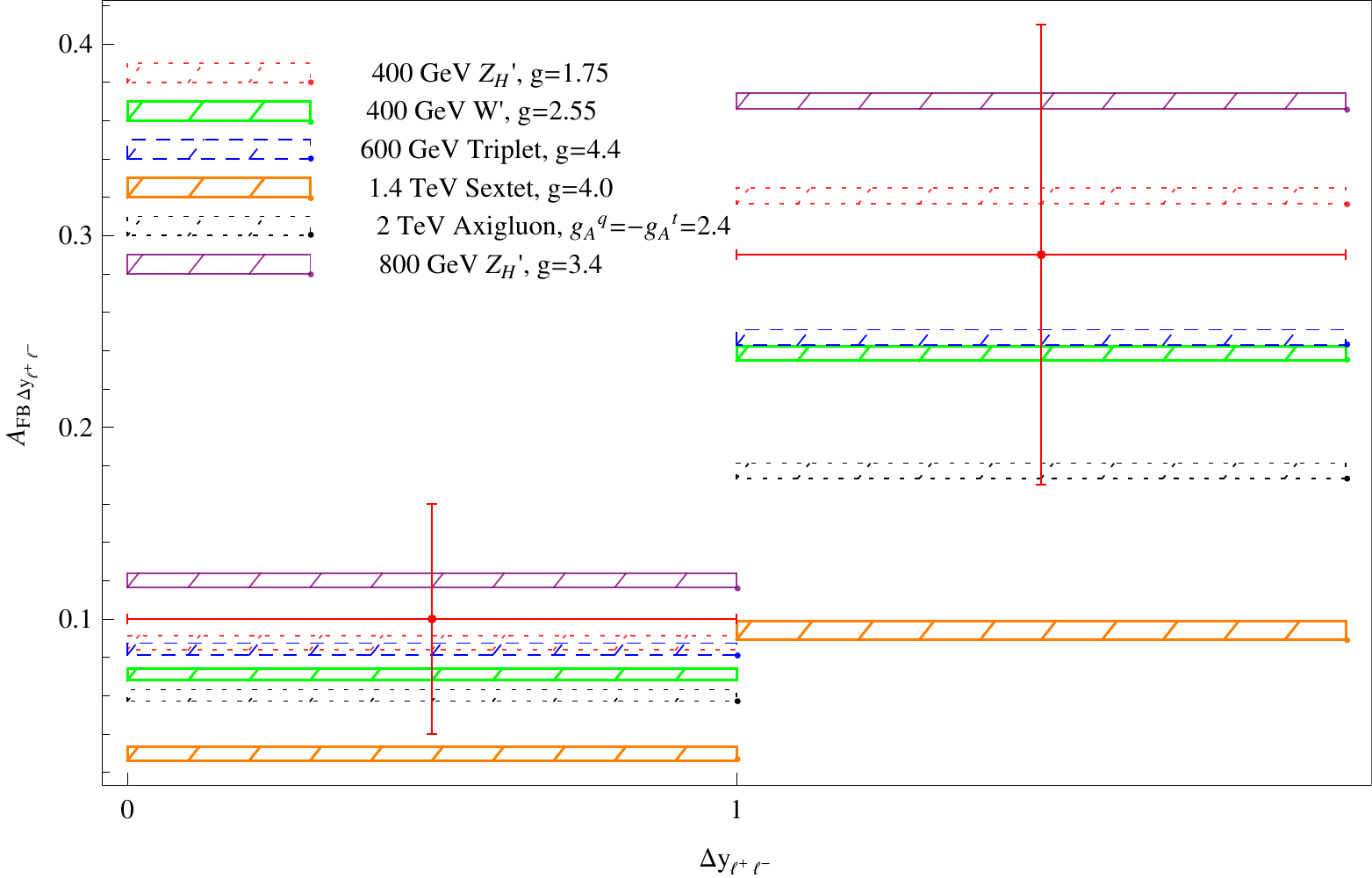} }
\subfigure[]{\includegraphics[width=0.45\textwidth]{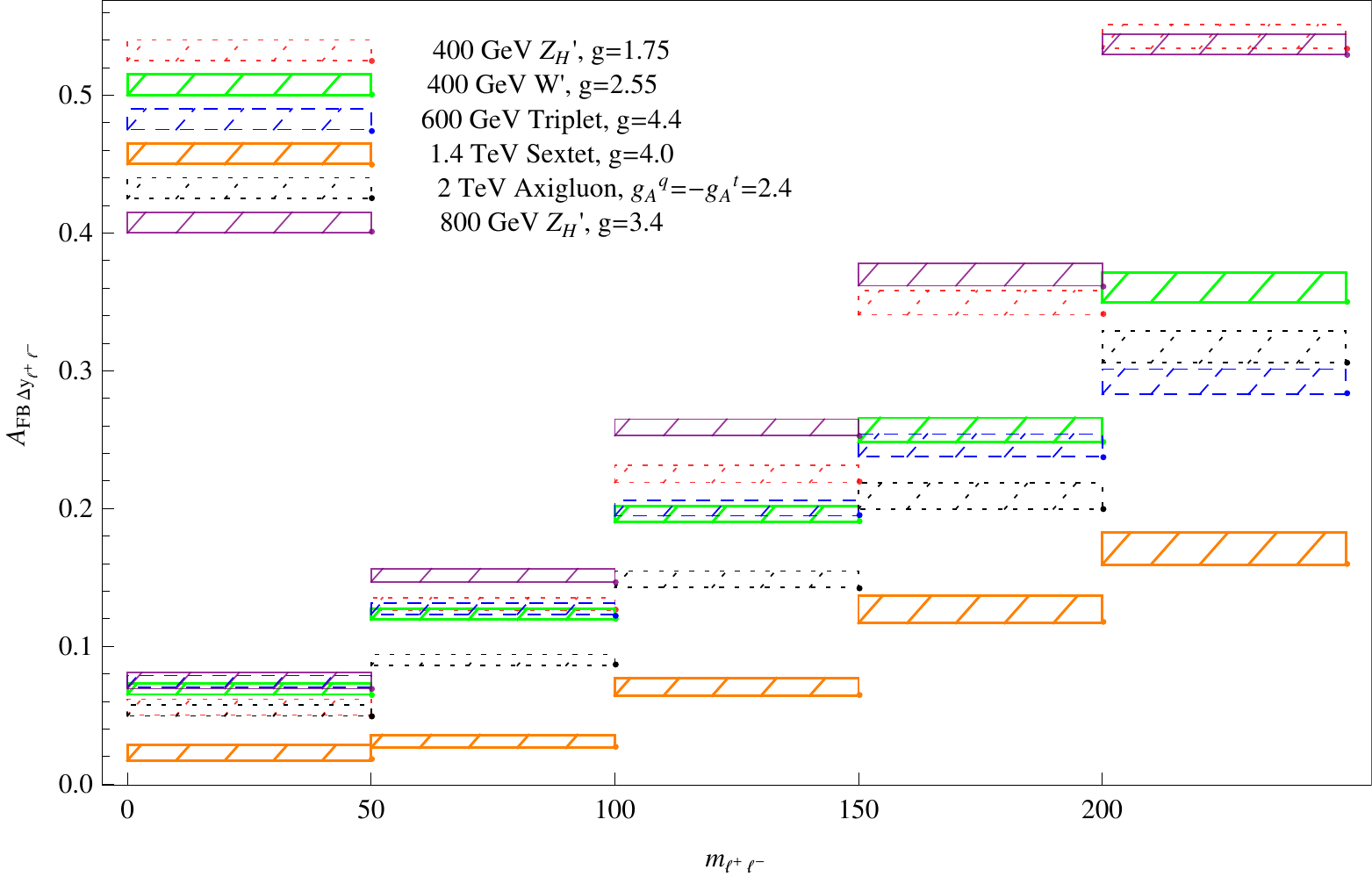}}
\end{center}
\caption[Parton level lepton forward-backward asymmetry.]{Parton level lepton forward-backward asymmetry  for $t \bar{t}$ events in which both tops decay leptonically, and in which both leptons pass the rapidity cuts corresponding to those used in the recent CDF analysis \cite{CDFLeptons}.  Red points are extracted from the results in \cite{CDFLeptons}.  {\bf (a): } $A_{FB (\Delta y_{\ell^+ \ell^-})} $ as a function of $|\Delta y| = |y_{\ell^+} - y_{\ell^-} | $.  {\bf (b):}  $A_{FB (\Delta y_{\ell^+ \ell^-})}$ $ = {N(y_{\ell^+} - y_{\ell^-} > 0) - N(y_{\ell^+} - y_{\ell^-} < 0) \over N(y_{\ell^+} - y_{\ell^-}> 0) + N(y_{\ell^+} - y_{\ell^-} < 0) } $ as a function of $\ell^+ \ell^-$ invariant mass.  }
\label{LepAFB}
\end{figure}

In order to avoid potential issues with the top reconstruction, one can also look at the asymmetries in di-leptons, where both tops decay leptonically.   The CDF collaboration recently reported results from an analysis of di-leptonic $t \bar{t}$ events. In addition to reporting asymmetries obtained after reconstructing tops in events, they report the raw lepton asymmetry \cite{CDFLeptons}.  We compare the raw lepton asymmetry in benchmark models at the showered parton level to the CDF measured value in Fig.~(\ref{LepAFB}a). To better compare with the CDF measurement, we show the asymmetry for only events that have electron rapidities in the range $|\eta| < 1.1$ or $1.2 < |\eta| < 2.8$ and muons with rapidities in the range $|\eta| < 1$, corresponding to the rapidity cuts placed on the leptons in their analysis.  We also point out that examining the lepton asymmetry as a function of lepton-lepton invariant mass could be instructive. We have shown the lepton forward-backward asymmetry as a function of $m_{\ell^+ \ell^-}$ in Fig.~(\ref{LepAFB}b).

\section{Early LHC Reach}

We now consider the feasibility of discovering such models in the early run of the LHC at 7 TeV.  To this end, we make use of our previous results \cite{Gresham:2011dg}.    According to these results, a 200 GeV $W'$ with coupling of 1 should be discoverable at $3 \sigma$ with 1 fb$^{-1}$ of data.  For a 400 GeV $W'$, the coupling must be larger than 1.3, and for a 600 GeV $W'$, the requirement is a coupling of 1.8.  Thus we see all of the $W'$ models giving rise to large asymmetries should be observable with an fb$^{-1}$ of data.  

Likewise, for the $Z'_H$ model, a coupling larger than 0.7 is required to discover a 200 GeV state at 3 $\sigma$ with 1 fb$^{-1}$, while a coupling of 0.8 is required for a 400 GeV state, and a coupling of 1.2 for a 600 GeV state.
For the triplets, a coupling of $\sim 0.9$ is required for a 3 $\sigma$ discovery with 1 fb$^{-1}$ for 400 GeV or lower masses; for a 600 GeV triplet, the requirement strengthens to requiring a coupling of 1.3.  Similar types of constraints can be obtained for the sextet models.  

The broad conclusion here is that all of the $t$-channel models that we considered here to fit the Tevatron top forward-backward asymmetry should give rise to $3\sigma$ excesses with 1 fb$^{-1}$ at the LHC in the context of a top-jet resonance search. 

According to the analysis in \cite{Bai:2011ed}, the axigluon benchmark models presented in this paper will be rapidly discoverable at the LHC through dijet events.

\section{Conclusions}

We examined models of new physics that could generate the top forward-backward asymmetry.  We considered $W'$, $Z'_H$ triplet and sextet diquarks, as well as axigluon models.  We compared the asymmetries produced by these models to those observed at the Tevatron, and concluded that of the models that generate a large enough asymmetry in the invariant mass bin $M_{t\bar{t}} > 450 \mbox{ GeV}$, the $Z'_H,~W'$ and axigluon models are the only ones that do not hugely overproduce the total $t\bar{t}$ production cross-section.  To bring the $W'$ models into agreement with the total $t\bar{t}$ production cross-section extracted at the Tevatron, we noted an important effect: the efficiency to select $t\bar{t}$ events from $W'$ models is significantly lower than for the Standard Model.  This same effect is also helpful in improving the agreement between the invariant mass spectra of the $W'$ and $Z'_H$ models with the Standard Model predictions (which agree with observations). Our result also differs from earlier studies on the diquark models which found that they could adequately produce the asymmetry without producing unduly large cross-sections.   In order to further investigate the model-dependence in the extracted asymmetry and invariant mass spectra, we then proceeded to decay the top quarks and simulate detector effects, reconstructing the tops via a likelihood based algorithm.  This allowed us to compare our results against the raw CDF results in the asymmetry as well as invariant mass spectra.  We found that when this was done, some $W'$ and $Z'_H$ models adequately reproduced the invariant mass spectra.  It also allowed us to compare our results against the CDF results for lab versus center-of-mass frames, as well as the 4 jet versus 5 jet asymmetries.  We conclude that while the models reproduce the observed decrease in the asymmetry in the 5 jet sample, no appreciable difference occurs between the lab and center-of-mass frames.  Lastly, we note that an LHC search at 7 TeV for top jet resonances could exclude at the $> 3 \sigma$ level any $Z'_H$, $W'$ or diquark model that produces the Tevatron asymmetry. 

{\em Acknowledgments}:  We thank Dan Amidei for many helpful conversations.

{\em Note added:}  While this work was being finalized, \cite{Ligeti:2011vt} appeared, which explores sextet and triplet models.  While our results agree quantitatively with theirs, our conclusions on the viability of these models for explaining the asymmetry  are more pessimistic, on account of the total $t\bar{t}$ cross-section and invariant mass distribution.


\end{document}